\documentclass[%
 reprint,
 amsmath,amssymb,
 aps,
 pre,
 nofootinbib
]{revtex4-2}

\usepackage{soul}
\usepackage{mathrsfs}
\usepackage{color}
\usepackage{amsfonts}
\usepackage{amssymb}
\usepackage{amsmath}
\usepackage{amsthm}
\usepackage{latexsym}
\usepackage{graphicx}
\usepackage{tabularx}
\usepackage{bm}
\usepackage{bbm}
\usepackage{array}
\usepackage{esvect}
\usepackage{import}
\usepackage{wrapfig}
\usepackage{csquotes}
\usepackage{dsfont}
\usepackage{xspace}
\usepackage[makeroom]{cancel}
\usepackage[font=footnotesize,labelfont=bf]{caption}
\usepackage{appendix}

\DeclareMathOperator{\arccosh}{arcosh}

\newcommand\ddfrac[2]{\frac{\displaystyle #1}{\displaystyle #2}}
\newcommand{\comment}[1]{}
\newcommand{\peclet}{P$\acute{\mathrm{e}}$clet\xspace}
\usepackage{scalerel}

\def\vecsign#1{\rule[1.388\LMex]{\dimexpr#1-2.5pt}{.36\LMpt}%
	\kern-6.0\LMpt\mathchar"017E}

\usepackage{stackengine,amsmath}
\stackMath

\begin{document}

\preprint{APS/123-QED}

\title{Cell shape and orientation control galvanotactic accuracy}

\author{Ifunanya Nwogbaga$^1$}
\author{Brian A. Camley$^{1,2}$}%
\affiliation{%
 $^1$Department of Biophysics, Johns Hopkins University, Baltimore, MD 21218\\
 $^2$Department of Physics \& Astronomy, Johns Hopkins University, Baltimore, MD 21218
}%

\begin{abstract}
Eukaryotic cells sense and follow electric fields during wound healing and embryogenesis -- this is called galvanotaxis. %
Galvanotaxis is believed to be driven by the redistribution of transmembrane proteins and other molecules, referred to as ``sensors", through electrophoresis and electroosmosis. 
Here, we update our previous model of the limits of galvanotaxis due to stochasticity of sensor movements to account for cell shape and orientation. Computing the Fisher information, we find that cells in principle possess more information about the electric field direction when their long axis is parallel to the field, but that for weak fields maximum-likelihood estimators of the field direction may actually have lower variability when the cell's long axis is perpendicular to the field. In an alternate possibility, we find that if cells instead estimate the field direction by taking the average of all the sensor locations as its directional cue (``vector sum''), this introduces a bias towards the short axis, an effect not present for isotropic cells. We also explore the possibility that cell elongation arises downstream of sensor redistribution. We argue that if sensors migrate to the cell's rear, the cell will expand perpendicular the field -- as is more commonly observed -- but if sensors migrate to the front, the cell will elongate parallel to the field.\\
\end{abstract}

\maketitle

\section*{\label{sec:intro}Introduction}
Electric fields play a pivotal role in biological systems. The pioneering experiments of Luigi Galvani in the 18th century on muscle contractions in frog legs  \cite{galvani1792viribus,galvani1954commentary,mccaig2005controlling} are some of the earliest examples showcasing their significance. Direct current (DC) electric fields with strengths ranging from 30--100 mV/mm have been measured within and around developing embryos \cite{shi1995three,hotary1994endogenous}, though applying additional electric fields to embryos cause abberations in embryonic development \cite{metcalf1994weak,hotary1992evidence}. Endogenous electric fields in the range of 40--200 mV/mm also emerge around wounds \cite{pullar2005cyclic,nuccitelli2003endogenous,nuccitelli2003role,funk2015endogenous}. Applying additional electric fields during wound healing accelerates the healing process \cite{ren2019keratinocyte,liang2020application,sari2019effect,li2020toward}. These electric fields have been shown to stimulate various types of cells to undergo galvanotaxis (alternately ``electrotaxis''), a phenomenon where cells migrate directionally in response to the electric field. Examples of galvanotaxing cells include keratocytes \cite{allen2013electrophoresis,kobylkevich2018reversing,zhu2016camp}, keratinocytes \cite{hart2013keratinocyte,pullar2005cyclic,trollinger2002calcium}, granulocytes \cite{gruler1990automatic,gruler1991neural,franke1990galvanotaxis}, fibroblasts \cite{gruler1986new,erickson1984embryonic,brown1994electric}, and neural crest cells \cite{gruler1991neural,nuccitelli1989extracellular,nuccitelli1993protein}. 
Some cells such as keratocytes and neural crest cells respond to field strengths as low as 25 mV/mm \cite{allen2013electrophoresis} and 10 mV/mm \cite{gruler1991neural}, respectively. 

There exists a consensus that cells sense electric fields through transmembrane molecule redistribution, which occurs via electrophoresis and electroosmosis. Simply put, proteins and other molecules on the cell surface are pulled by the electric field and migrate relative to the field, accumulating on one side of the cell. %
This redistribution of membrane-bound components is crucial for the initiation of galvanotaxis \cite{kobylkevich2018reversing,allen2013electrophoresis,mclaughlin1981role,nwogbaga2023physical,brown1994electric,sarkar2019electromigration}. Putative sensor candidates identified or hypothesized in the past include EGFR, P$_2$Y , integrins, and lipid rafts \cite{riding2016atp,fang1999epidermal,huang2009involvement,pullar2006beta4,lin2017lipid}. Different cell types, of course, may have different sensors, and there may be more than one sensor. Ion channels have also been observed to play a role \cite{trollinger2002calcium,yang2013epithelial}, suggesting there may be multiple mechanisms for sensing the field orientation \cite{lasota2024dynamics}. 

In our previous manuscript \cite{nwogbaga2023physical}, we determined the physical limits of galvanotactic measurement through sensor redistribution for round cells (circular and spherical). We developed a model that quantified how cells measure the direction of the electric field by using maximum likelihood estimation (MLE). {In our approach, we assume that cells sense electric fields solely through redistribution of their transmembrane sensors. We discovered that circular cells can predict the field's direction efficiently by employing a simple strategy: the cell follows the direction of the average position of the sensors on its surface. We call this strategy ``vector sum". The same result holds in 3D for spherical cells. This result comes naturally out of MLE -- for round cells, MLE and vector sum produce the same measurement, though we find in this work this is not always true.} In this manuscript, we extend the work of \cite{nwogbaga2023physical} to investigate how the shape and orientation of an elliptical cell influence its estimate of the field direction. %

\begin{figure}[ht]
	\centering
	\includegraphics[width=\linewidth]{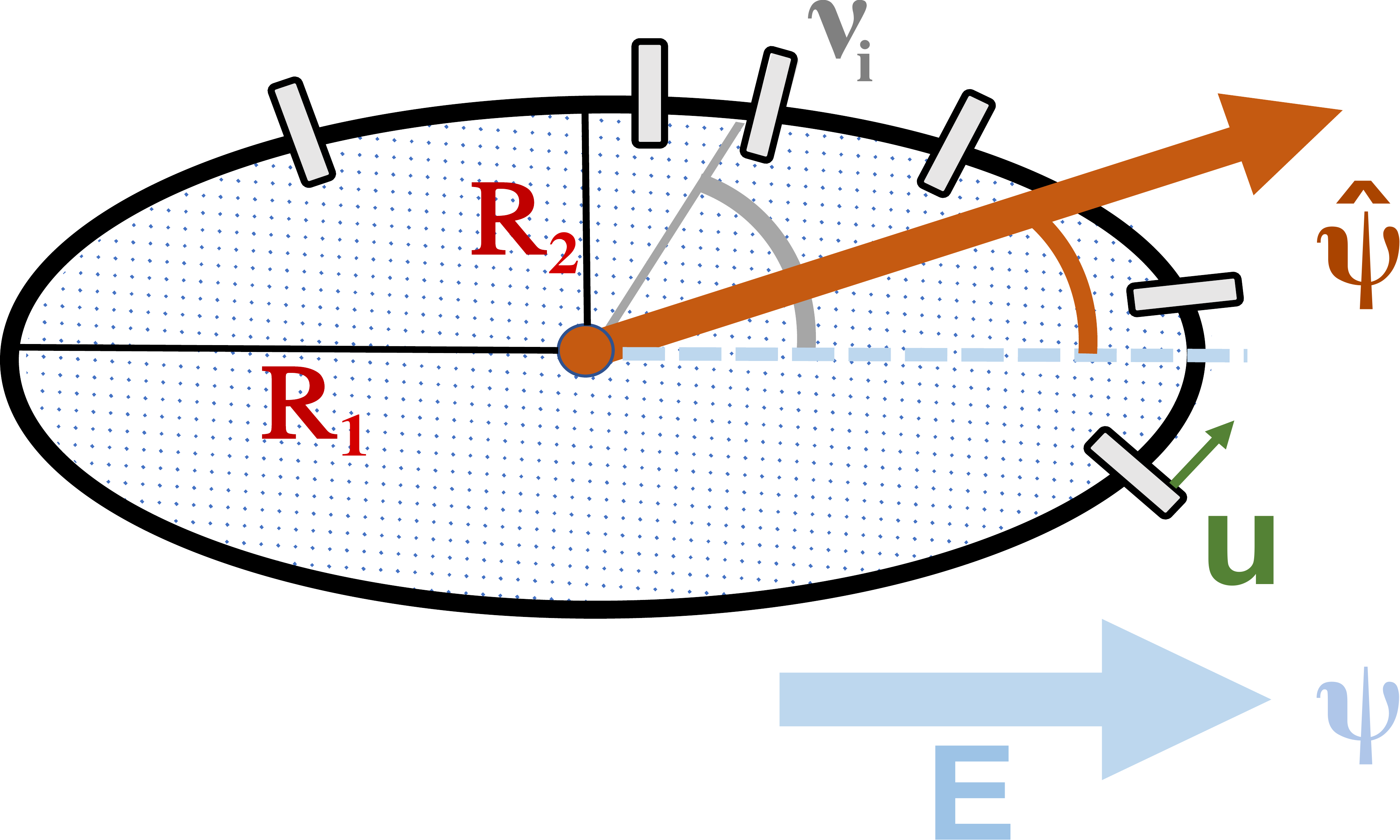}
	\caption{Illustration of sensors on the surface of an elliptical cell; sensors are labeled by their elliptic angle location $\nu_i$ and traveling with velocity $\mathfrak{u}$. The cell is in an electric field pointing along the angle $\psi$ and measures an estimated angle $\hat\psi$.}
	\label{fig:schematicfield}
\end{figure}

We want to find the factors limiting the accuracy of a cell's estimate $\hat\psi$ of the angle of the applied field $\psi$. To do this, we calculate the electric field around the cell, which controls the velocity of the sensors. 
We use this field to find the concentration of sensors around the cell arising from the competition between diffusion and electromigration. We construct a probability distribution of the sensor location from the concentration, and use maximum likelihood and an extended Cram\'er-Rao bound to quantify the cell's estimate of the electric field direction and its variance \cite{nwogbaga2023physical}. Surprisingly, whether the cell is more accurate in sensing the field's orientation when its long axis is parallel to or perpendicular to the field direction depends on the strength of the field and the method the cell uses to interpret the sensor location. Under the most likely experimental conditions, cells are better sensors of field orientation when the field is perpendicular to the cell's long axis, as in chemotaxis \cite{hu2011geometry}. However, the difference in accuracy between the best and worst orientations may not be large if the cell uses a maximum-likelihood estimate of the field.  If the cell instead uses a simpler-to-compute ``vector sum'' estimate, its estimate will be biased, but can in some circumstances be quite close to the maximum-likelihood estimate, while at stronger fields, the cell can be quite far from the best possible estimate, and its accuracy will depend much more strongly on cell orientation relative to the field. We also address the possibility that the cell can change shape in response to an applied field, hypothesizing that cells extend radially proportional the level of sensor concentration in a graded radial extension-style model \cite{lee1993principles,camley2017crawling,ohta2009deformationreaction}. We find this predicts that the cell would lengthen along the field direction if the sensors are at the front of the cell, contrary to known observations \cite{shim2021overriding}, but lengthen perpendicular to the field if the sensors are at the back of the cell. These results suggest a potential way to determine the sign of the sensor mobility -- something our earlier work did not constrain. %

\begin{figure*}[htbp]
	\centering
	\includegraphics[width=\linewidth]{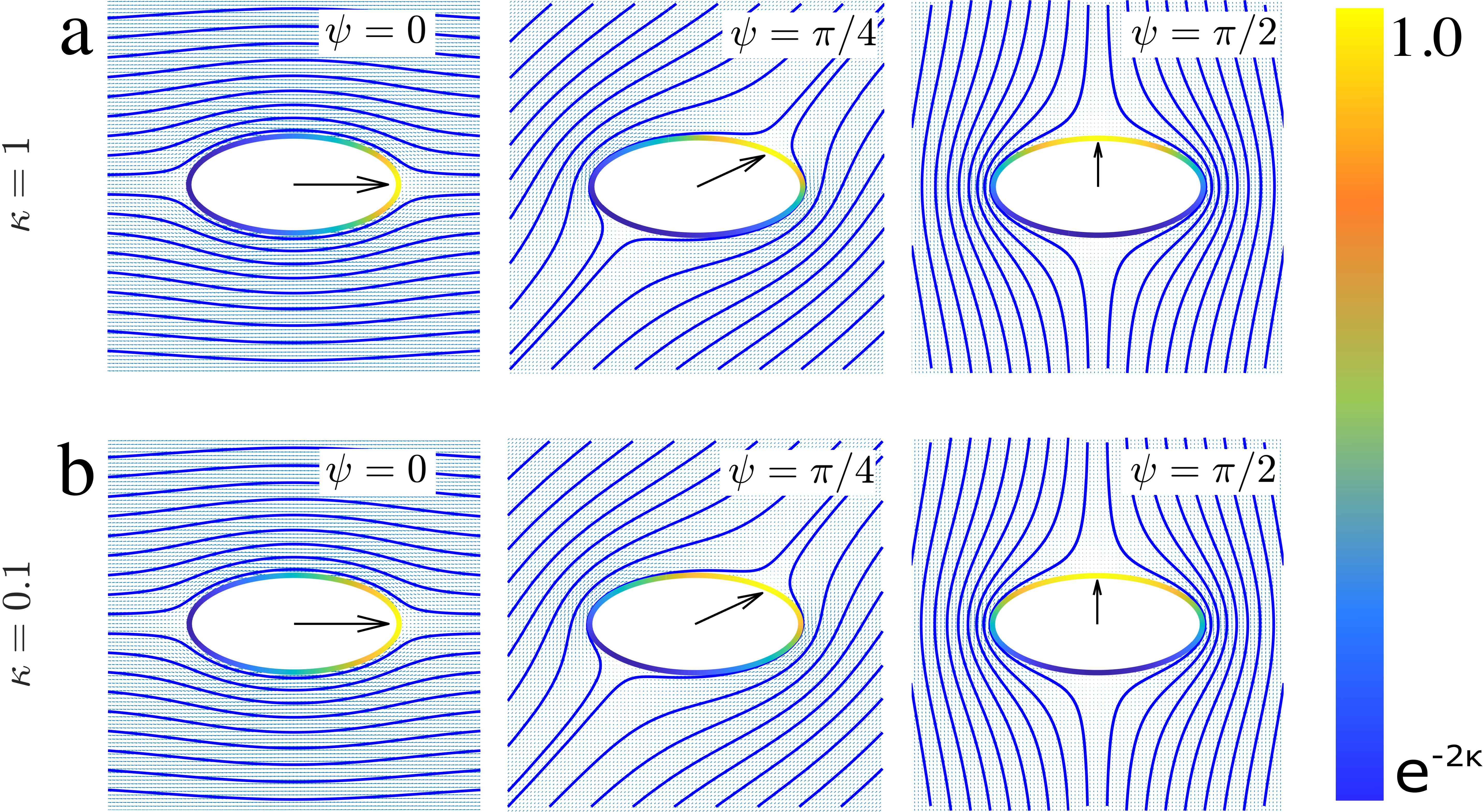}
	\caption{Field lines of the field outside the cell, given three different orientations $\psi$ of the applied field. Note that from the boundary conditions (Appendix \ref{app:boundary}), the normal component of the electric field vanishes at the cell boundary. Sensor concentration $c(\nu)$ is also plotted as a color map. The concentration is plotted as $c(\nu)/c_\mathrm{max}$, where $c_\mathrm{max}=c_0 e^\kappa$. This normalization sets the maximum of the colorbar to 1 and the minimum to $e^{-2\kappa}$. (a) Field lines and sensor concentration when $\kappa=1$. (b) Field lines and sensor concentration when $\kappa=0.1$. }
	\label{fig:fieldlines}
\end{figure*}

\section*{\label{sec:model}Model and Methods}
We describe our cell as a two-dimensional ellipse, characterized by major and minor axes $R_1$ and $R_2$. $\lambda = R_1/R_2$ is the cell's aspect ratio. Throughout this paper, we will choose coordinates so that the cell is elongated along the $x$-direction -- the field direction may vary but the cell's long axis direction is always consistent. Motivated by \cite{allen2013electrophoresis,nwogbaga2023physical,sarkar2019electromigration}, we claim that the cell senses the electric field direction $\psi$ via electromobile sensors (Fig. \ref{fig:schematicfield}), producing an estimate of the field direction $\hat{\psi}$. Given the geometry, we use elliptic coordinates $(\mu,\nu)$, where $\mu$ is the elliptic radius and $\nu$ is the elliptic angle, with corresponding unit vectors $\bm{\hat{\mu}}(\mu,\nu)$ and $\bm{\hat{\nu}}(\mu,\nu)$ (see Appendix \ref{app:ellipticcoordinates}). {$\bm{\hat{\mu}}$ is normal to the cell surface while $\bm{\hat{\nu}}$ is tangential to the surface}.

\subsection*{Calculation of the external electric field}
The motion of sensors on the cell boundary is dependent on the local electric field \cite{nwogbaga2023physical,mclaughlin1981role,kobylkevich2018reversing}. To find the electric field at the boundary of the cell, we solve Laplace's equation  $\nabla^2\Phi=0$ for the electric potential $\Phi$ in elliptic coordinates, subject to the conditions that far away from the cell $\Phi=-E_0(x\cos\psi+y\sin\psi)$, and at the cell boundary the gradient of the potential vanishes normal to the cell surface, i.e. the electric field is solely tangential to the membrane. The latter condition is valid for a nonconductive membrane \cite{kotnik2000analytical,mossop2004electric}. This assumption is appropriate since the resistivity of the membrane is much higher than that of the solution and cytoplasm (see Appendix \ref{app:boundary}). Subsequently, the electric field around the cell is solely determined by the cell shape \cite{kotnik2000analytical,mossop2004electric} (Fig. \ref{fig:fieldlines}). %
The solution to Laplace's equation in elliptic coordinates is well known \cite{morse1954methods,moon2012field}. %
When an external field  %
$\mathbf{E}_\mathrm{ext}=E_0(\cos \psi,\sin \psi)$ is applied to the cell, the electric field outside the cell is
\begin{equation}
	\begin{split}
		\mathbf{E}=\dfrac{E_0a}{h_\mu}\dfrac{e^\mu-e^{2\mu_0-\mu}}{2}\cos(\nu-\psi)\bm{\hat{\mu}}\\
		-\dfrac{E_0a}{h_\mu}\dfrac{e^\mu+e^{2\mu_0-\mu}}{2}\sin(\nu-\psi)\bm{\hat{\nu}}.
	\end{split}
\end{equation}
The cell has foci at $\pm a$, with $a=\sqrt{R_1-R_2}$ (see Appendix \ref{app:ellipticcoordinates}); the cell's boundary is at $\mu=\mu_0$. %
The scale factor $h_\mu(\mu,\nu)=a\sqrt{(\cosh2\mu-\cos2\nu)/2}$. This is the usual curvilinear scale factor for elliptic coordinates, $h_\mu=\left\vert\partial\mathbf{r}/\partial\mu\right\vert$. The field tangent to the cell at the boundary $\mu=\mu_0$ is proportional to the unit vector $\bm{\hat{\nu}}$, which points in the direction tangential to the cell surface
\begin{equation}
	\label{eq:tangential}
	\mathbf{E}_\parallel=-E_0\dfrac{ae^{\mu_0}}{h_{\mu_0}}\sin(\nu-\psi)\bm{\hat{\nu}},
\end{equation}
where the scale factor $h_{\mu_0}\equiv h_\mu(\mu_0,\nu)$. Note that $ae^{\mu_0}=a\cosh\mu_0+a\sinh\mu_0=R_1+R_2$. The largest magnitudes of $\mathbf{E}_\parallel$ are naturally where the applied field is parallel to the local surface of the cell. This means that when the applied field is parallel to the long axis of the cell ($\psi = 0$), the local tangential field $\mathbf{E}_\parallel$ is largest along the long side of the cell, tending to concentrate sensors to the narrow tips of the cell. By contrast, if the field is perpendicular to the cell, the tangential field is largest at the narrow tips of the cell, and the sensors are pulled toward the broad side of the cell, though they are not as strongly concentrated. Plots of $\mathbf{E}_\parallel$ are shown in  Appendix \ref{app:electricfieldellipse}. %

\subsection*{Transport of sensors along boundary and probability distribution}

Armed with an expression for the electric field everywhere around the cell, we can now calculate the transport of sensors at the cell boundary. The sensors move following the tangential field to the surface, traveling along the cell's boundary with velocity $\mathfrak{u}$. We assume that $\mathfrak{u}=\mathfrak{m}\mathbf{E}_{\parallel}$, where $\mathfrak{m}$ is the electrophoretic mobility constant. This mobility constant can be positive or negative depending on the sensors' charge and the zeta potential of the cell surface \cite{mclaughlin1981role}; we do not attempt to estimate $\mathfrak{m}$, but think of it as phenomenological and generally try to fit it from experimental data. The assumption that the velocity is proportional to the local electric field is reasonable since the cell's membrane is treated as nonconducting and the electric double layer is thin compared to the radius of curvature of the cell \cite{mclaughlin1981role,overbeek1967interpretation}. Each sensor is labeled by its elliptic angle location $\nu_i$. Transport of the density of sensors $c$ along the contour can be derived by balancing the advective flux $c \mathfrak{u}$ with the diffusive flux. This flux balance captures the competition between sensor migration driven by the electric field, which tends to polarize the sensors, and diffusion, which tends to spread the sensors out \cite{mclaughlin1981role,nwogbaga2023physical}. The transport of sensors is explained by the equation
\begin{equation}
	\label{eq:flux}
	c \mathfrak{u}=D\nabla_\parallel c,
\end{equation}
where $\nabla_\parallel$ is the gradient along the membrane and $D$ the sensor's diffusion coefficient. The density of sensors can only depend on their location on the boundary $\nu$, so we can write Eq. \eqref{eq:flux} explicitly as:
\begin{equation}
	c(\nu)\mathfrak{m} \mathbf{E}_{\parallel}=D\dfrac{1}{h_{\mu_0}}\dfrac{\partial c(\nu)}{\partial\nu}\bm{\hat{\nu}}.
\end{equation}
The scale factor $h_{\mu}$ appears in this calculation from the expression of the gradient in the  $\bm{\hat{\nu}}$ direction in elliptic coordinates. Substituting in Eq. \eqref{eq:tangential} for $\mathbf{E_\parallel}$ and rearranging, the solution is proportional to the von Mises distribution \cite{mardia1975algorithm}
\begin{equation}
	c(\nu)=c_0\exp{[\kappa\cos{(\nu-\psi)}]}. \label{eq:concentration}
\end{equation}
Here, $c_0$ is a proportionality constant set to determine the total number of sensors. $\kappa = \mathfrak{m}E_0 ae^{\mu_0}/D =\mathfrak{m}E_0 (R_1+R_2)/D$ is the ratio between the rate of electromigration and diffusion (\peclet number \cite{leal2007advanced}). If $\kappa$ is larger, electromigration dominates diffusion, and the sensors are more concentrated (note the extent of the yellow region in Fig. \ref{fig:fieldlines}a versus b). We often want to think about $\kappa$ as a ratio between the strength of the applied electric field $E_0$ and some characteristic field scale, $1/\beta$, so we write $\kappa=\beta E_0$, where $\beta = \mathfrak{m}(R_1+R_2)/D$. %
For circular cells with radius $R_0$ \cite{nwogbaga2023physical}, we found a similar result of concentration proportional to $e^{\beta E_0 \cos (\theta-\psi)}$ in polar coordinates, with $\beta =\mu R_0/D$. (Note that $\mu$ here is not the elliptic radius!) Our result in Eq. \eqref{eq:concentration} will more obviously limit back to this result if we define $\beta=\bar{\mathfrak{m}}\bar{R}/D$, where $\bar{\mathfrak{m}}=2\mathfrak{m}$ and $\bar{R}=(R_1+R_2)/2$. $c(\nu)$ is plotted in Fig. \ref{fig:fieldlines}a--b in conjunction with the electric field lines.

We note that in our prior manuscript \cite{nwogbaga2023physical} we assumed $\mathfrak{u} = \mu \mathbf{E}_{0,\parallel}$, where $\mathbf{E}_0$ was the {\it applied} field and $\mu$ is a mobility constant. In this manuscript there is a subtle difference. Now, $\mathfrak{u} = \mathfrak{m}\mathbf{E}_\parallel$, where $\mathbf{E}_\parallel$ is the {\it local} electric field tangent to the cell boundary and $\mathfrak{m}$ is the mobility constant. For circular cells, either choice gives the same answer, but with a conventional factor of 2 between $\mu$ and $\mathfrak{m}$ (see Appendix A of \cite{nwogbaga2023physical}), but for elliptical cells, this is not the case. Therefore, we have chosen the more physically-realistic assumption $\mathfrak{u} = \mathfrak{m}\mathbf{E}_\parallel$.

The concentration of Eq. \eqref{eq:concentration} gives the number of sensors per unit length of boundary; we will need to work with $\mathcal{P}(\nu)$, the probability density per elliptic angle $\nu$. Changing variables, this can be found as
\begin{equation}
	\label{eq:probability}
	\mathcal{P}(\nu)=Z^{-1}c(\nu)h_{\mu_0},
\end{equation}
where 
\begin{equation}
	Z=\int_{\psi-\pi}^{\psi+\pi} c(\nu)h_{\mu_0}\mathrm{d}\nu. \label{eq:Z}
\end{equation}
(See Fig. \ref{fig:probdist} and Appendix \ref{app:probdist} for more details.) %

\subsection*{Maximum likelihood estimation and Fisher information}

Following \cite{nwogbaga2023physical}, we use maximum likelihood estimation to quantify the cell's estimate of the field's direction.  We ask ourselves ``What electric field direction maximizes the likelihood of observing a given probability distribution of sensors?" The likelihood function for $N$ noninteracting sensors, each at elliptic angle $\nu_i$, is given by 
\begin{equation}
	\mathcal{L}(E,\psi;\{\nu\})=\prod_{i=1}^{N}\mathcal{P}(\nu_i),
\end{equation}
however, the log-likelihood is easier to work with

\begin{equation}
	\label{eq:loglikelihood}
	\ln{\mathcal{L}}=\sum_{i=1}^{N}\left[\kappa\cos{(\nu_i-\psi)}+\ln{h^{(i)}_{\mu_0}}\right]-N\ln{Z}.
\end{equation}
Here, $h^{(i)}_{\mu_0}=a\sqrt{\left(\cosh2\mu_0-\cos2\nu_i\right)/2}$, where the superscript $i$ acts as a label for sensor $i$. To minimize $\ln \mathcal{L}$, we differentiate, setting $\partial_\psi\ln{\mathcal{L}}|_{\hat{\psi}}=0$ to get an expression that gives us the estimator $\hat\psi$
\begin{equation}
	\label{eq:MLE}
	\dfrac{1}{N}\sum_{i=1}^{N}\kappa\sin{(\nu_i-\hat{\psi})}=\dfrac{1}{Z}\dfrac{\partial Z}{\partial\psi}.
\end{equation}
The integrals that define $Z$ and $\partial_\psi Z$ on the right hand side of Eq. \eqref{eq:MLE} have to be evaluated numerically in general, but have an analytical solution in the limit of near-circular cells (see Appendix \ref{app:MLE_FI_pert}). {In the special case of a circular cell, the elliptic angle $\nu_i$ tends to the polar angle $\theta_i$, the right hand side of Eq. \eqref{eq:MLE} equals zero, and the cell's prediction $\hat\psi$ of the true field angle $\psi$ can be solved explicitly by the equation \cite{nwogbaga2023physical}
\begin{equation}
    \label{eq:MLE_ReceptorSum}
	\tan{\hat\psi}=\dfrac{\sum_i\sin{\theta_i}}{\sum_i\cos{\theta_i}} \; \; \; \textrm{(circular cells)}.
\end{equation}}
The Fisher information can also be derived from the log-likelihood by calculating $\mathcal{I}=\langle-\partial^2_\psi\ln\mathcal{L}\rangle$ (see Appendix \ref{app:MLE_FI} for derivation):
\begin{equation}
	\label{eq:FisherInformation}
	\mathcal{I}=N\kappa^2\left[\left\langle\sin^2{(\nu-\psi)}\right\rangle-\left\langle\sin{(\nu-\psi)}\right\rangle^2\right].
\end{equation}
We can compute the averages in Eq. \eqref{eq:FisherInformation} numerically, $\left\langle \cdots\right\rangle=\int \cdots \mathcal{P}(\nu)\mathrm{d}\nu$. The Fisher information reveals how much information the cell would have about the field direction given a particular distribution of sensors. The Fisher information is crucial for quantifying the variability in the cell's estimate. Since the estimate $\hat\psi$ is periodic in nature, we characterize variability in the estimate with the circular variance $2(1-\langle\cos{(\hat\psi-\psi)}\rangle)$, which will reduce to the ordinary variance in the limit where the deviation between $\hat\psi$ and $\psi$ is small. The circular variance of $\hat\psi$ has a lower bound given by an extension of the Cram\'er-Rao bound (derived in Appendix D of \cite{nwogbaga2023physical})
\begin{equation}
	\label{eq:circularvariance}
	2(1-\langle\cos{(\hat\psi-\psi)}\rangle)\geq2\left(1-\sqrt{\dfrac{\mathcal{I}}{1+\mathcal{I}}}\right)
\end{equation}
or, equivalently, $\langle\cos{(\hat\psi-\psi)}\rangle\leq\sqrt{\mathcal{I}/(1+\mathcal{I})}$. The lower bound for the circular variance becomes more accurate as number of sensors increases or the sensors become more polarized  -- in this limit, the inequality becomes an equality. We will compute the Fisher information $\mathcal{I}$ for an elliptical cell in order to determine the accuracy with which it can sense the field angle $\psi$. 

\subsection*{Stochastic simulations}
In addition to our analytic and numerical results, we also conduct stochastic simulations using randomly-generated configurations of sensors. We generate a configuration of $N$ sensors for each of $N_\mathrm{cell}$ cells. The configurations were generated by the rejection sampling method, drawing $N$ sensor positions independently from Eq. \eqref{eq:probability} \cite{casella2004generalized}. Briefly, we propose a uniformly-distributed sensor location $\nu_i\sim\mathcal{U}([\psi-\pi,\psi+\pi])$, as well as $u_i\sim\mathcal{U}([0,1])$. Then, we compute $p_i=\mathcal{P}(\nu_i)/\mathrm{sup}\{\mathcal{P}(\nu)\}$. If $p_i>u_i$, then the proposed $\nu_i$ is selected as one of the sensor positions. Otherwise the proposal is rejected. This process is repeated until $N$ sensors are generated. 

If we assume the cells are estimating the field by maximizing the likelihood, the cell's estimate $\hat{\psi}$ is calculated by numerically finding the $\hat{\psi}$ that maximizes the log-likelihood, using  Eq. \eqref{eq:loglikelihood}. We use the Nelder-Mead simplex algorithm (MATLAB's fminsearch) \cite{lagarias1998convergence}. We will also sometimes assume cells choose a direction as the vector sum of unit normals pointing from the sensor locations (Eq. \eqref{eq:discreteaverageRS}). 

During our simulations (for both MLE and vector sum), we fix the the average of the cell semi-major and minor axes as $\bar{R}=20~\mu$m (typical keratocyte radius from Fig. 4a of \cite{keren2008mechanism}). If $\lambda$ is fixed, it is fixed at $\lambda=3$, the reasonable upper limit for aspect ratio of a keratocyte; see Fig. 4b of \cite{keren2008mechanism}). $R_2$ was selected so $(R_1+R_2)/2=(1+\lambda)R_2/2=\bar{R}$ remains fixed (see Appendix \ref{app:ellipticcoordinates}).

\section*{\label{sec:results}Results}

\subsection*{Accuracy of galvanotaxis depends on cell orientation, aspect ratio, and \peclet number $\kappa$}

\begin{figure}[bp]
	\centering
	\includegraphics[width=\linewidth]{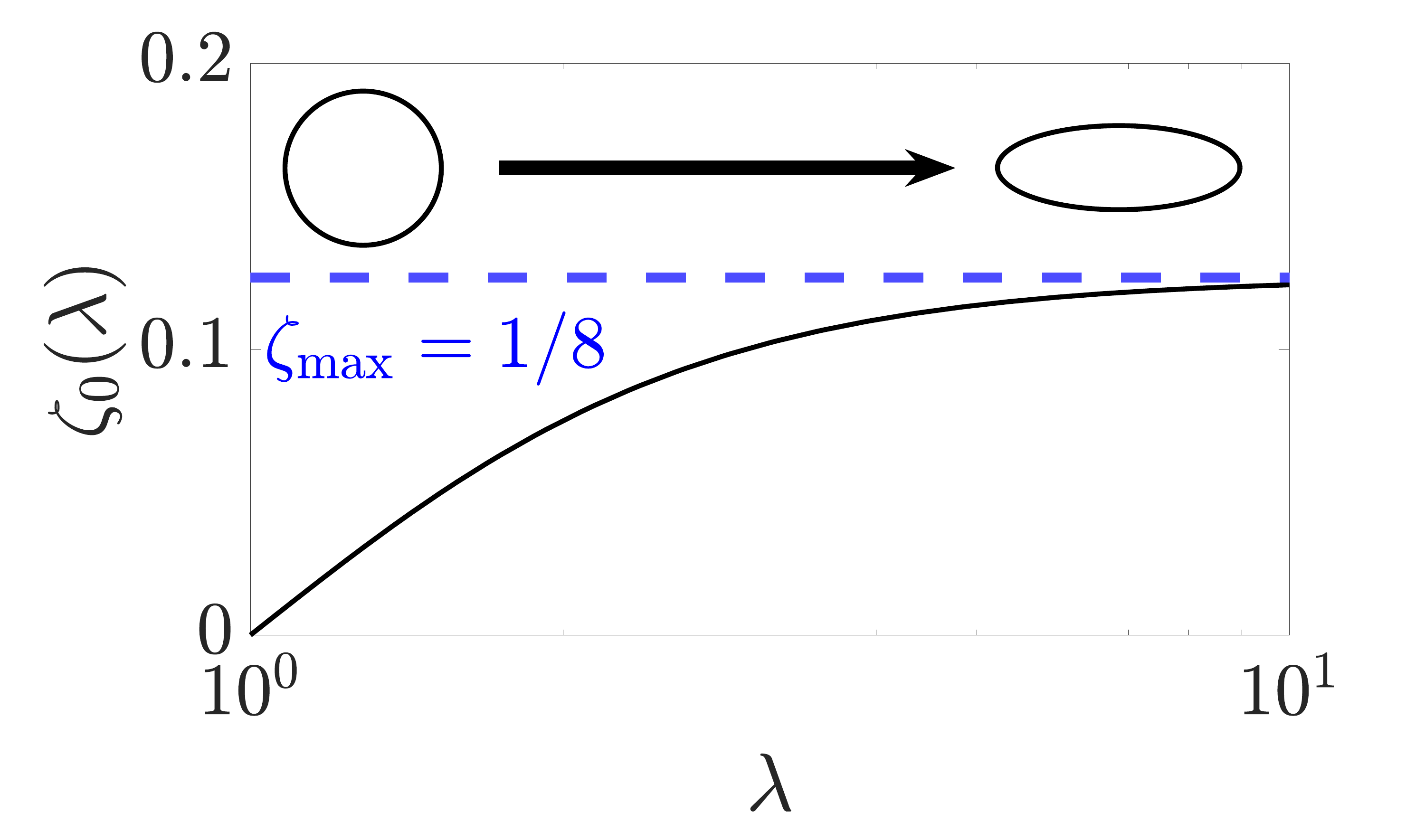}
	\caption{Aeolotropic constant $\zeta_0$ as a function of aspect ratio $\lambda$; the constant increases and saturates as a cell becomes more elliptical, approaching its maximum of 1/8.}
	\label{fig:zeta}
\end{figure}

\begin{figure*}[ht]
	\centering
	\includegraphics[width=\textwidth]{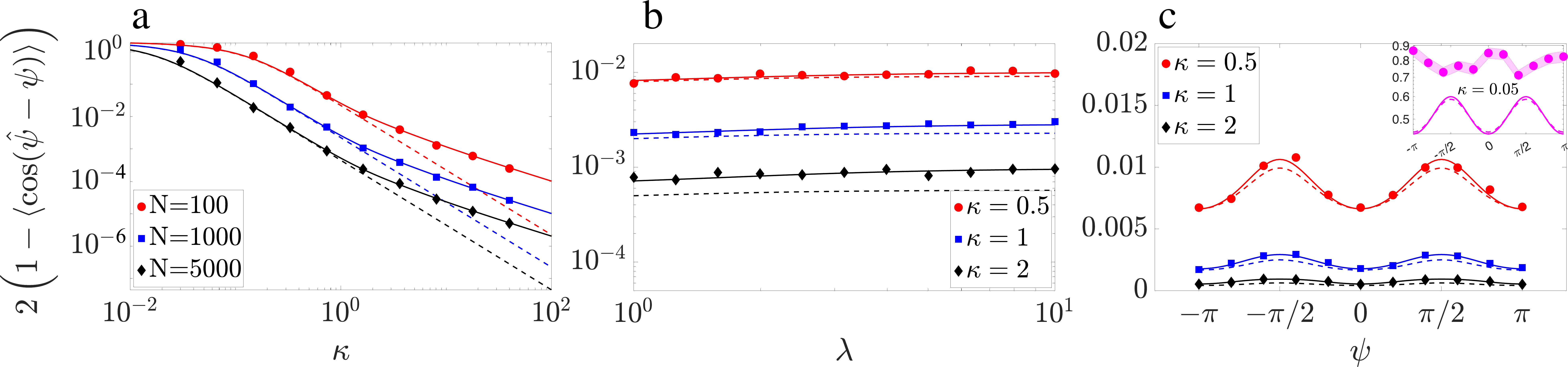}
	\caption{Circular variance changes as a function of the \peclet number $\kappa$, aspect ratio $\lambda$, field direction $\psi$, and number of sensors $N$. Solid lines are given by Eq. \eqref{eq:circularvariance}, which was calculated by numerically computing the Fisher information of Eq. \eqref{eq:FisherInformation} and inserting it into Eq. \eqref{eq:circularvariance}. Dashed lines are the analytic approximation of Eq. \eqref{eq:FI_simplifyweak}. Symbols are stochastic simulation averaged over 500 cells (a) or 1000 cells (b, c). Error bars (standard error of the mean) are on the order of the symbol size. (a): Circular variance as a function of the \peclet number $\kappa$, plotted for various values of number of sensors $N$. The aspect ratio $\lambda=3$. The field angle relative to the long axis of the cell is $\psi=\pi/3$. (b): Circular variance as a function of the aspect ratio $\lambda$, plotted for various values of $\kappa$. The number of sensors is $N=1000$. The field angle is $\psi=\pi/3$. (c): Circular variance as a function of the true field direction $\psi$, plotted for various values of $\kappa$. The number of sensors is $N=1000$. The aspect ratio is $\lambda=3$. The inset plots same thing as (c) but for $\kappa=0.05$. Here, the shaded region around the simulation symbols represent the error bars, which is the standard error of the mean.}
	\label{fig:circularvariance}
\end{figure*}
The Fisher information $\mathcal{I}$ tells us how much information the cell has about the electric field direction from its sensor distribution. It can be hard to draw intuition from Eq. \eqref{eq:FisherInformation} as it must be evaluated numerically. Fortunately, in the limit of nearly circular cells with sensors whose motion is dominated by diffusion (small $\kappa$), the Fisher information $\mathcal{I}$ simplifies to (see Appendices \ref{app:MLE_FI_pert}-\ref{app:FI_approx}) 
\begin{equation}
	\label{eq:FI_simplifyweak}
	\mathcal{I}= N\kappa^2\left(\dfrac{1}{2}+\zeta_0\cos2\psi\right).
\end{equation}
The aeolotropic constant $\zeta_0$ characterizes the anisotropic contribution to the Fisher information. In other words, it accounts for the cell's deviation from a circle. It is a function of the aspect ratio $\lambda$,
\begin{equation}
	\zeta_0(\lambda)=\dfrac{1}{8}\dfrac{\lambda^2-1}{\lambda^2+1}.
\end{equation}
$\zeta_0$ is positive for all aspect ratios $\lambda \ge 1$, and $\zeta_0$ goes to zero for circular cells and $\zeta_\mathrm{max} \to 1/8$ for infinitely elliptical cells e.g. one-dimensional lines  (Fig. \ref{fig:zeta}). 

Eq. \eqref{eq:FI_simplifyweak} predicts that the cell's amount of information varies depending on its orientation with respect to the field, with the Fisher information $\mathcal{I}$ maximal when the field is along the major axis and minimal when the field is along the minor axis. From the perspective of the model, this is not surprising. We see in Fig. \ref{fig:fieldlines} that the sensors are more concentrated when the cell's long axis is parallel to the field. Therefore we expect more information about the cell's orientation in this configuration. How much information does the cell have when the field is parallel to the cell ($\psi = 0$) versus perpendicular to the cell ($\psi = \pi/2$)?  %
Because $\zeta_0$ is at most 1/8, we can see from Eq. \eqref{eq:FI_simplifyweak} that the largest the ratio between the maximum information at $\psi = 0$ and minimum at $\psi = \pi/2$ is $\mathfrak{R}_\mathrm{max}=5/3$. This corresponds to an at most  $\sim67\%$ increase in information in the ``best'' orientation versus the ``worst'' orientation. We note that this result depends on Eq. \eqref{eq:FI_simplifyweak}, which is only valid in the limit of near-circular cells, where $\lambda$ is not too far from unity. However, we see in Fig. \ref{fig:circularvariance}b that Eq. \eqref{eq:FI_simplifyweak} is a suitable approximation for the full solution in Eq. \eqref{eq:FisherInformation} even out to $\lambda \approx 10$ if $\kappa \approx 0.5$. So for any reasonable cell shape, we are confident in Eq. \eqref{eq:FI_simplifyweak} as long as $\kappa$ is sufficiently small.

The Fisher information, combined with the bound of Eq. \eqref{eq:circularvariance}, tells us the minimum possible circular variance for an unbiased estimate of the direction $\psi$. We plot in Fig. \ref{fig:circularvariance}a how the circular variance $2(1-\langle\cos{(\hat\psi-\psi)}\rangle)$ varies with the \peclet number $\kappa$. Unsurprisingly, we see that $2(1-\langle\cos{(\hat\psi-\psi)}\rangle)$ decreases with increasing $\kappa$ and increasing number of sensors $N$. In Fig. \ref{fig:circularvariance}, we show the bound of \eqref{eq:circularvariance} using both the analytical solution for the weak-field Fisher information approximation Eq. \eqref{eq:FI_simplifyweak} (dashed line), and a full numerical solution to the Fisher information (solid line), and stochastic simulations (symbols). The small-$\kappa$ approximation deviates from the numerical solution and stochastic simulations at large $\kappa$, as expected.

We see in Fig. \ref{fig:circularvariance}a that the simulations generally agree with the full numerical solution (solid lines) -- but there is a small systematic deviation at very low $\kappa$ ($\kappa \ll 1$). In this regime, the simulations yield slightly higher values than what the theory predicts. This behavior is expected, and similar to what we found in \cite{nwogbaga2023physical}; see that paper's Fig. 2. Eq. \eqref{eq:circularvariance} is a lower bound for the circular variance, so it is always possible that the circular variance of the maximum likelihood estimator is above this bound. We found previously that for circular cells \cite{nwogbaga2023physical}, the circular variance of the maximum likelihood estimator converges onto our modified Cramer-Rao bound in the limit of large amount of information (large $\kappa$ and large $N$). (This convergence is guaranteed because that the maximum likelihood estimator is asymptotically efficient \cite{kay1993fundamentals}.)

In Fig. \ref{fig:circularvariance}b, we see that circular variance $2(1-\langle\cos{(\hat\psi-\psi)}\rangle)$ changes only slightly as aspect ratio increases while holding $\kappa$ constant. Changing aspect ratio $\lambda$ while holding $\kappa$ constant means that $(R_1+R_2)/2$ is held constant while $\lambda=R_1/R_2$ is changing, meaning that as a cell simultaneously becomes longer and thinner. This sort of change leads to only small differences in circular variance. The sign of the change of circular variance with $\lambda$ will depends on $\psi$, since the Fisher information can increase or decrease with aspect ratio depending on the sign of $\cos2\psi$ (Eq. \eqref{eq:FI_simplifyweak}); we have plotted Fig. \ref{fig:circularvariance}b with the field at angle $\psi = \pi/3$, and the long axis of the cell, as always, along the $x$-direction. %

Fig. \ref{fig:circularvariance}c reveals that the error in the cell's measurement depends on its orientation relative to the field, as we expected from the discussion of Eq. \eqref{eq:FI_simplifyweak}. We see that the circular variance is minimal when the field is pointed along the cell's major axis ($\psi = 0$) where there is the most information, and maximal along the minor axis where there is the least information ($\psi = \pi/2$). All of the plots in the main body of Fig. \ref{fig:circularvariance}c, which are in the range $\kappa \ge 0.5$, show good agreement between simulations and the bound of Eq. \eqref{eq:circularvariance} (symbols near-exactly overlapping with the solid line). However, as we mentioned above discussing Fig. \ref{fig:circularvariance}a, we know that the bound is no longer tight in the limit of smaller $\kappa$ -- the circular variance will exceed the bound. We plot this limit in the inset of Fig. \ref{fig:circularvariance}c. Surprisingly, not only does the circular variance of the maximum-likelihood estimate of the simulations exceed the bound, but the maximal circular variance occurs at $\psi = 0$, where the bound is minimal. This means that, in this limit of very small fields, cells with their long axis perpendicular to the field will have lower error in measuring the field angle. The origin of this effect is not clear, and we explore it further below.

\begin{figure*}[tp]
	\centering
	\includegraphics[width=\textwidth]{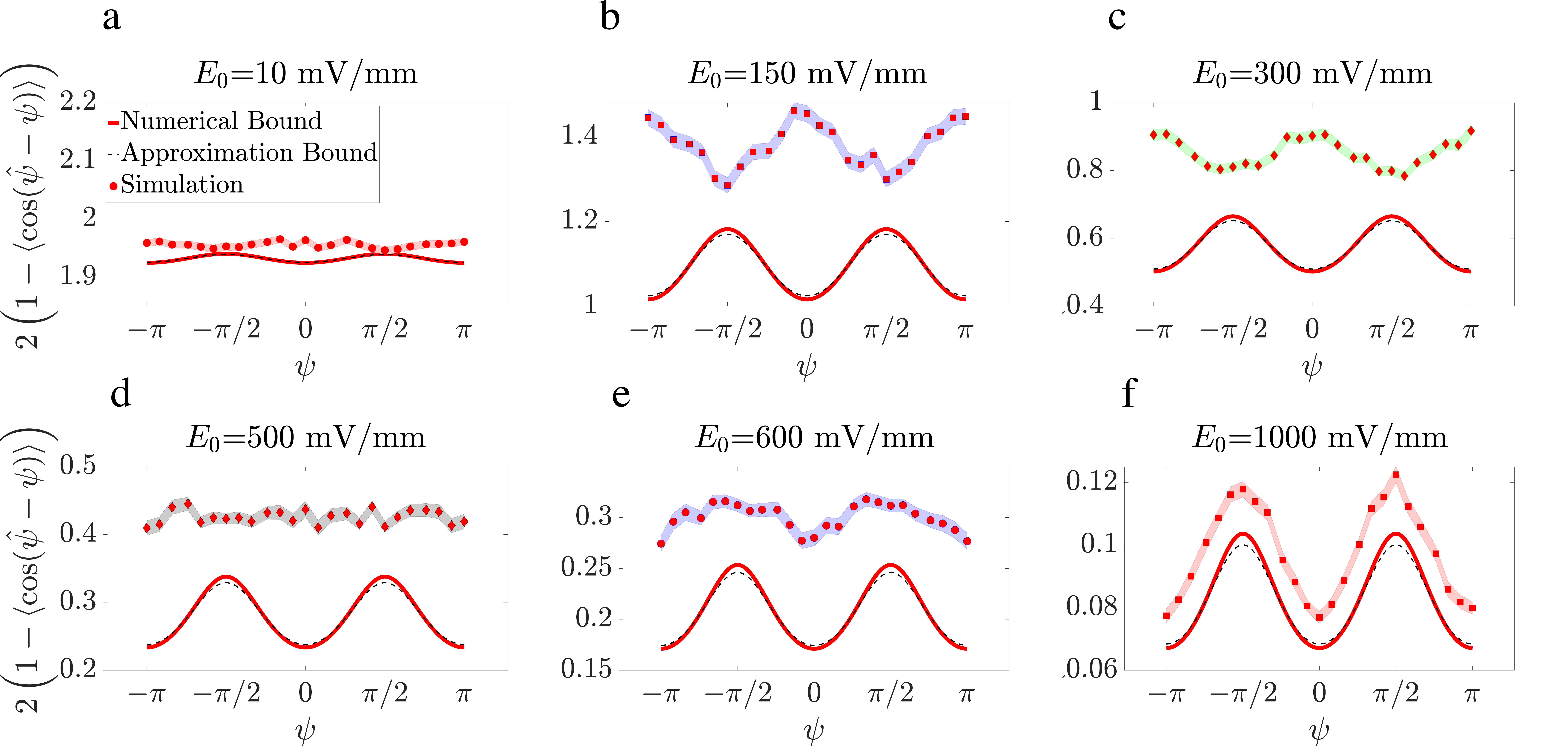}
	\caption{Circular variance as a function of the field orientation $\psi$ for keratocytes using MLE. The aspect ratio $\lambda=3$, $\gamma=3.4\times10^{-3}$ mm/mV, and the number of sensors $N=1000$ are kept constant. Solid red lines are given by Eq. \eqref{eq:circularvariance}, which was calculated by numerically determining Fisher information of \eqref{eq:FisherInformation}, and plugging back into Eq. \eqref{eq:circularvariance}. Dashed black lines are using the analytic approximation Eq. \eqref{eq:FI_simplifyweak}. Symbols are stochastic simulation averaged over 5000 cells (\ref{fig:keratocyte}a is averaged over 50000 cells). Shaded regions around the symbols represent the standard error of the mean.}
	\label{fig:keratocyte}
\end{figure*}

\subsection*{Modeling keratocytes as ellipses can minimize semi-minor axis variance perpendicular to the electric field under weaker field strengths}

In Fig. \ref{fig:circularvariance}c, we have seen that, depending on $\kappa$, cells' best orientation could be {\it either} with their long axis parallel to the field or perpendicular to the field. Which of these limits is relevant for real cells? We will fit our model to the case of fish keratocytes, a well-studied model system for cell locomotion and galvanotaxis \cite{allen2013electrophoresis, allen2020cell, keren2008mechanism, lee1993principles, lee1993fish}.
We fit our model, assuming the weak-field limit of Eq. \eqref{eq:FI_simplifyweak}, to data measuring keratocyte directionality as a function of electric field strength  \cite{sun2013keratocyte} (details in Appendix \ref{app:directionality}). Directionality measures the average angle between the cell's velocity and the applied electric field; we think of the cell's velocity as a proxy for its estimate of the field direction, so the directionality is just $\langle\cos{(\hat\psi-\psi)}\rangle$. We can then fit the directionality data to get the parameters of our model by using the approximation $\langle\cos{(\hat\psi-\psi)}\rangle\approx e^{-\mathcal{I}^{-1}/2}$ (Appendix \ref{app:directionality}). We note that our formula for $\mathcal{I}$ does not separately depend on $N$ and $\kappa$ but only on their product $N\kappa^2 = N \beta^2 E_0^2$, so we cannot separately fit $N$ and $\beta$. We define, as in \cite{nwogbaga2023physical}, $\gamma^2=N\beta^2/2$. $1/\gamma$ is a characteristic field strength. Then, Eq. \eqref{eq:FI_simplifyweak} can be written as $\mathcal{I} = \gamma^2E_0^2(1 + 2\zeta_0\cos2\psi)$. We choose $\lambda = 3$ as a rough estimate for keratocytes \cite{keren2008mechanism}, which sets the aeolotropic constant $\zeta_0(3)=0.1$. As keratocytes typically migrate with their long axis aligned perpendicular to the electric field \cite{allen2013electrophoresis, allen2020cell}, $\psi = \pi/2$, and the Fisher information for keratocytes can be simplified to 
\begin{equation}
\mathcal{I}=0.8\gamma^2E_0^2 \; \; \textrm{(keratocytes} \perp \textrm{to field}). \label{eq:keratocyte_FI}
\end{equation}
If we fit experimentally-measured directionality as a function of the electric field using Eq. \eqref{eq:keratocyte_FI} and $\langle\cos{(\hat\psi-\psi)}\rangle\approx e^{-\mathcal{I}^{-1}/2}$ (see Appendix \ref{app:directionality} and Fig. \ref{fig:keratocytefitdirectionality}), we find $\gamma=3.4\times10^{-3}$ mm/mV. 

Do keratocytes have a preferred orientation in which they sense more accurately? Fig. \ref{fig:keratocyte}b plots the circular variance as a function of field angle using our keratocyte estimate for $\gamma$ at a typical field strength of $E_0=150$ mV/mm. We see that while the bound given by Eq. \eqref{eq:circularvariance} predicts the minimal variance at $\psi = 0$, stochastic simulations predict the opposite trend -- the minimal circular variance is at $\psi = \pi/2$. (This result does not depend on the number of sensors $N$ as long as we hold the magnitude of the Fisher information constant by keeping $\gamma^2 = N\beta^2/2$ fixed; see Fig. \ref{fig:keratocytefit} in Appendix \ref{app:keratocytefit}.) To illustrate where this transition in preferred axis happens, we plot the circular variance at $\gamma=3.4\times10^{-3}$ mm/mV for increasing field strengths in Fig. \ref{fig:keratocyte}a-f. We can see a wide variety of trends, ranging from cells all having near-uniformly distributed estimators (circular variance $\approx 2$) at $E_0 = 10$ mV/mm to cells having minimal error when perpendicular to the field ($E_0 = 150, 300$ mV/mm) transitioning to a flat dependence on angle ($E_0 = 500$ mV/mm) and eventually seeing the trend predicted by the Fisher information at larger field ranges (1000 mV/mm).  This range of field strengths is the experimentally relevant range. It goes from the weakest field strengths that cells types such as neural crest cells have responded to ($E_0=10$ mV/mm \cite{gruler1991neural}) to the strongest fields typically used in galvanotaxis experiments ($E_0=1000$ mV/mm \cite{sarkar2019electromigration}). Note the the difference between the peaks and troughs of the circular variance lower bound and simulations become smallest at low fields ($E_0 = 10$ mV/mm), high fields ($E_0 = 1000$ mV/mm), and the transition ($E_0 = 500$ mV/mm).

The result of Fig. \ref{fig:keratocyte} is somewhat dramatic. The trend in circular variance of the maximum likelihood estimate can be the {\it opposite} of the bound predicted by the Fisher information. This is surprising! We certainly initially expected that even if the circular variance exceeded the bound, it would follow a similar trend. We do not have a clear explanation for why this occurs. One possibility is that this behavior arises due to some odd property of the maximum likelihood estimator (or our numerical evaluation of the estimator) -- but we will see in the next section that a more physically plausible estimator also has lower variance when the field is perpendicular to the cell's long axis.

\subsection*{Vector sum is a biased but plausible sensing strategy}

\begin{figure*}[tp]
	\centering
	\includegraphics[width=\textwidth]{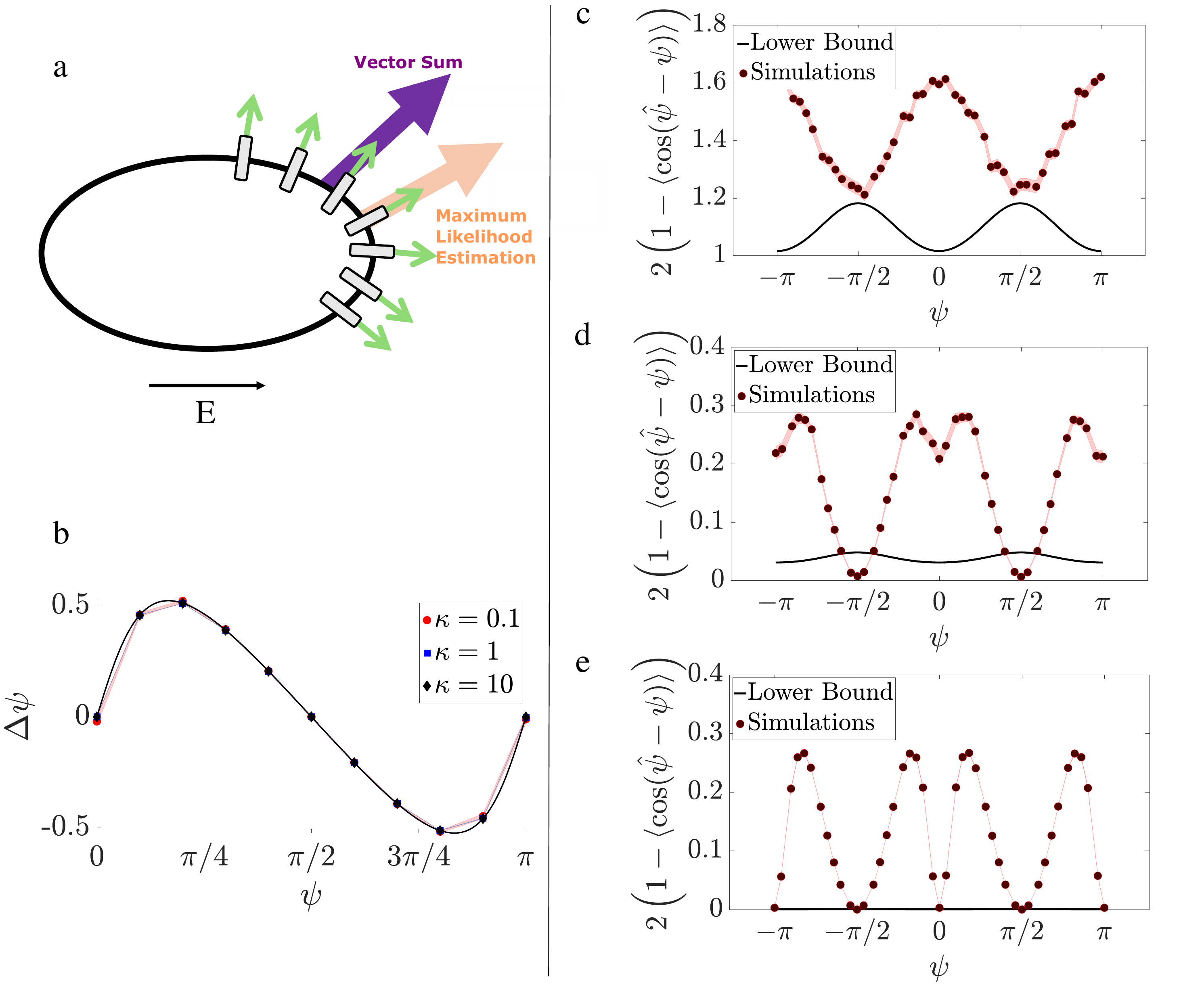}
	\caption{(a): Schematic showcasing the bias that vector sum introduces relative to MLE. (b): Vector sum as a function of field angle. Symbols are simulation, averaged over 500 cells. Shaded area between the symbols indicates standard error of the mean. Solid lines are numerical calculation of Eq. \eqref{eq:averageRS}. Here, we have $N=10000$ and $\lambda=3$. $\langle \hat\psi_\mathrm{VS} \rangle$ is computed here by first computing $\hat\psi_\mathrm{VS}$ for each cell and then averaging. (c): Circular variance plot for keratocytes calculated using vector sum. $N=10000$, $\gamma=3.4\times10^{-3}$ mm/mV, and $\kappa\approx5\times10^{-3}$. Black line is the lower bound from Eq. \eqref{eq:circularvariance}. Simulations are represented by the symbols and are averaged over 5000 cells. Shaded region are standard error of the mean. (d): Circular variance plot for keratocytes calculated using vector sum. $N=10000$, $\gamma=3.4\times10^{-3}$ mm/mV, and $\kappa\approx5\times10^{-2}$. Black line is the lower bound from Eq. \eqref{eq:circularvariance}. Note that the simulations drop below the lower bound, as the lower bound assumes an {\it unbiased} estimator, which is not true for the vector sum. Simulations are represented by the symbols and are averaged over 500 cells. Shaded region are standard error of the mean. (e): Circular variance plot for keratocytes calculated using vector sum. $N=10000$, $\gamma=3.4\times10^{-3}$ mm/mV, and $\kappa\approx5\times10^{-1}$. Black line is the lower bound from Eq. \eqref{eq:circularvariance}. Simulations are represented by the symbols and are averaged over 500 cells. Shaded region indicates standard error of the mean.}
	\label{fig:receptorsum}
\end{figure*}

We have used the maximum likelihood estimation approach in our earlier results because it is asymptotically efficient. In the limit of large numbers of observations, it should reach the Cramer-Rao bound. However, as a model for a biological organism, it has the clear downside of being extremely complicated to compute since cells presumably do not have Nelder-Mead optimization easily available. We found in \cite{nwogbaga2023physical}  that the maximum likelihood estimator for a {\it circular} cell is the direction of the vector sum of vectors pointing to each sensor. This estimate could be easily computed if the cell exerts a protrusive force localized at each sensor, pointing normally outward (see Appendix K of \cite{nwogbaga2023physical}). This scheme is physically plausible, and consistent with the idea that local charge regulates protrusion \cite{banerjee2022spatiotemporal}. In this section, we study the properties of cells that estimate directions with vector sum, taking the sum of the unit normals $\hat{\bm{\mu}}_i$ for each sensor $i$ pointing out from the ellipse,
\begin{align}
\label{eq:discreteaverageRS}
\bar{\bm{\mu}}&=\dfrac{1}{N}\sum_i^N\hat{\bm{\mu}}_i,\\[10pt]
\hat{\psi}_\mathrm{VS}
\label{eq;psiRS}&=\arctan{(\bar{\mu}_y/\bar{\mu}_x)}.
\end{align}

Unlike our results for circular cells, for an ellipse, the vector sum estimator $\hat{\psi}_\mathrm{VS}$ and the maximum-likelihood estimate $\hat{\psi}_\textrm{MLE}$ are not identical. In fact, we see in Fig. \ref{fig:receptorsum}a and \ref{fig:receptorsum}b that the vector sum estimate is biased.
 In Fig. \ref{fig:receptorsum}b, we plot the difference between the vector sum estimate, averaged over 500 simulated cells, and the true field orientation, $\Delta\psi=\langle\hat{\psi}_\mathrm{VS}\rangle-\psi$ as a function of the true field orientation. %
 We find that the bias in the vector sum estimator vanishes, $\Delta \psi = 0$, if the field is pointing along the long axis ($\psi = 0, \pi$) or the short axis ($\psi = \pi/2, 3\pi/2$), and is a maximum in between these orientations.  %
We see that $\langle\hat{\psi}_\mathrm{VS}\rangle$ does not point precisely in the true direction $\psi$ but has a consistent bias toward the minor axis. This bias does not appear to depend on $\kappa$ (Fig. \ref{fig:receptorsum}b). In addition to the simulations, we also calculate $\hat{\psi}_\mathrm{VS}$ from the average of the unit normal,
\begin{equation}
	\label{eq:averageRS}
	\langle\hat{\bm{\mu}}\rangle=\int_{\psi-\pi}^{\psi+\pi}\hat{\bm{\mu}}\mathcal{P}(\nu)\mathrm{d}\nu,
\end{equation}
and, in a slight abuse of notation, $\hat{\psi}_\mathrm{VS}=\arctan{(\langle{\hat{\mu}_y}\rangle/\langle\hat{\mu}_x\rangle)}$. We can compute this integral analytically, finding that $\langle \hat{\mu}_x \rangle = \mathcal{A}(\kappa) \sinh \mu_0 \cos \psi$ and $\langle \hat{\mu}_y \rangle = \mathcal{A}(\kappa) \cosh \mu_0 \sin \psi$ with $\mathcal{A}(\kappa)=2\pi a I_1(\kappa)Z^{-1}$, which is a function that depends on $\kappa$ and cell shape. $I_1(\kappa)$ is a modified Bessel function of the first kind. Since the aspect ratio of the cell is $\lambda = \cosh \mu_0 / \sinh \mu_0$ (Appendix \ref{app:ellipticcoordinates}), we see that the vector sum points in the direction $(\cos \psi, \lambda \sin \psi)$. This means that the vector sum becomes more biased to be parallel to the short axis for more elongated cells.  Our formula also shows that the bias is solely geometrical. It does not depend on $\kappa$ (Fig. \ref{fig:receptorsum}b). (A minor technical note: our finding that $\Delta\psi \neq 0$ does not, strictly speaking, tell you that the estimator must be biased in a periodic sense \cite{nwogbaga2023physical,mardia2000directional}. However, we do also find that $\langle \sin (\hat\psi_\textrm{VS}-\psi) \rangle$ is nonzero.)

How accurate is the vector sum? If cells compute using the vector sum, are they better at measuring when parallel to the field or perpendicular to the field? We use parameters appropriate to the keratocytes, as we did in Fig. \ref{fig:keratocyte}, and compute the circular variance as a function of the field angle (Fig. \ref{fig:receptorsum}c). We find that vector sum also predicts that the cell has the smallest variance when the long axis is perpendicular to the field ($\psi = \pm \pi/2)$. In fact, for $\psi = \pm \pi/2$, the cell is nearing our bound Eq. \eqref{eq:circularvariance} (Fig. \ref{fig:receptorsum}c). However, we note that the vector sum is not constrained by this bound -- it could give circular variances below the bound, since we derived Eq. \eqref{eq:circularvariance} under the assumption of an unbiased estimator. %
When $\kappa$ is increased from the physiologically relevant value of $\kappa \approx 5 \times 10^{-3}$ (Fig. \ref{fig:receptorsum}c) by a factor of 10 (Fig. \ref{fig:receptorsum}d), this corresponds to increasing the field strength from 150 mV/mm to 1500 mV/mm. However, even at this high electric field, the variance of the vector sum is minimal when the cell's long axis is perpendicular the field, though there is a dip in the circular variance curve near $\psi = 0$. This should be contrasted to Fig. \ref{fig:keratocyte}, in which a similar change in magnitude of electric field completely inverts the trend. We also note that in Fig. \ref{fig:receptorsum}d, the vector sum circular variance may fall below the bound, as is expected given the biased estimation -- our extended Cram\'er-Rao bound \cite{nwogbaga2023physical} was derived under the assumption of an unbiased estimator $\hat\psi$.

In Fig. \ref{fig:receptorsum}e, $\kappa$ is increased further by two orders of magnitude from the case of Fig. \ref{fig:receptorsum}c. Though this limit is likely no longer experimentally relevant, we see that increasing $\kappa$ magnifies the difference in sensing accuracy between having a cell with one of it principal axis aligned with the field versus not having one aligned. In this limit, the circular variance is at a minimum when the field is either along the long {\it or} the short axis of the cell. We believe this reflects the bias in the vector sum strategy. When $\kappa$ becomes large, the error bound from stochasticity shrinks relative to the size of the bias, and the systematic errors of order $\sim 0.5$ radians shown in Fig. \ref{fig:receptorsum}b are large relative to the variability from stochastic sensor locations. When this bias is relevant, as in Fig. \ref{fig:receptorsum}e, there are strong consequences for not sensing the field along one of the principal axes, as the error can be orders of magnitude higher off of any main axis.

\subsection*{Cell shape expands parallel or perpendicular to the field, depending on sensor mobility}

\begin{figure*}[tp]
	\centering
	\includegraphics[width=\textwidth]{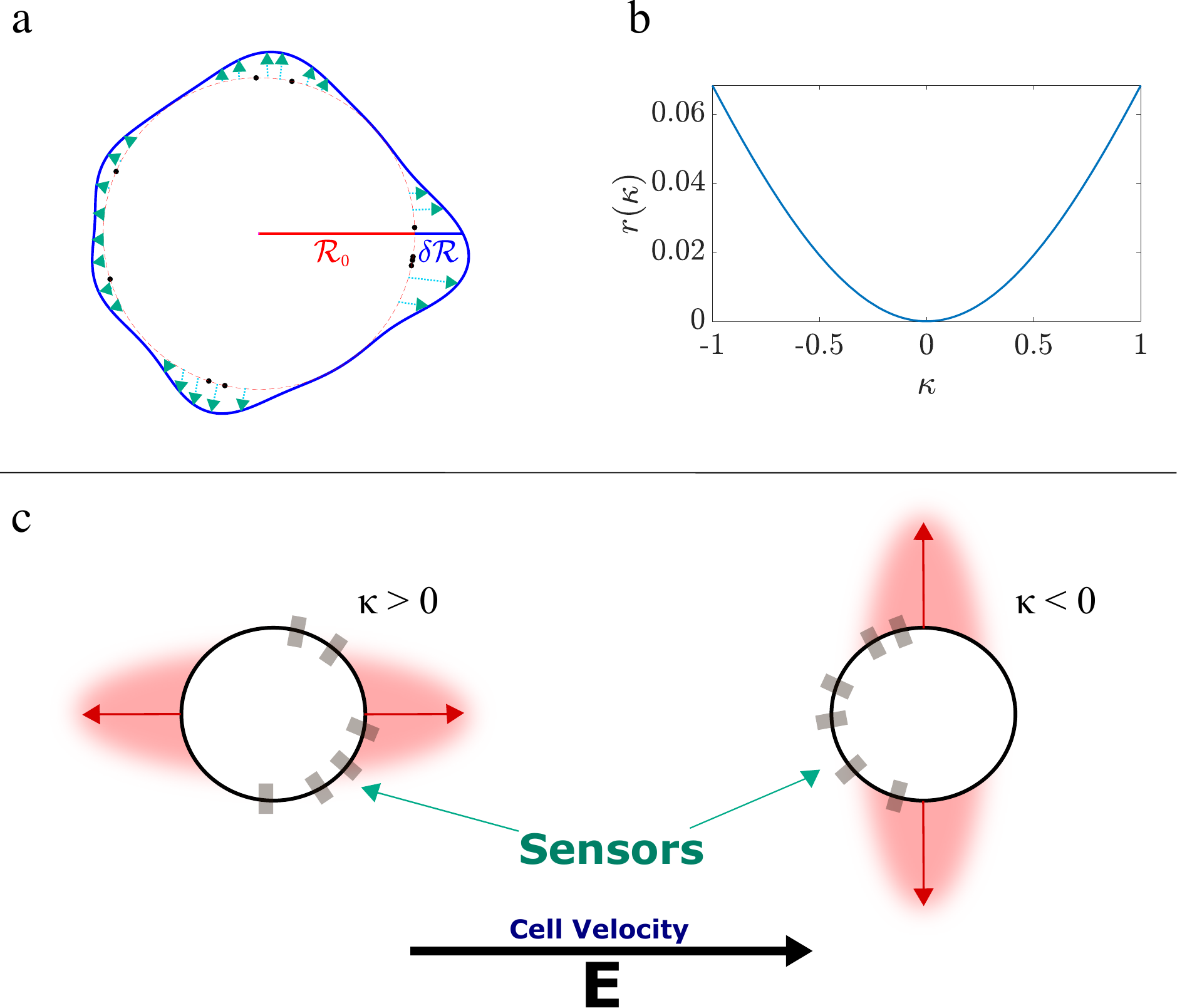}
	\caption{(a): Schematic of cell whose protrusions occur at the sensor locations (black points). The radius $\mathcal{R}=\mathcal{R}_0+\delta\mathcal{R}$. (b): Plot of $\mathfrak{r}(\kappa)$ showing that it is nonnegative for both positive and negative $\kappa$. (c): Cell with either sensors transported parallel to the applied electric field $(\kappa>0)$, causing the cell to expand parallel to the field, or with sensors transported opposite to the applied electric field $(\kappa<0)$, causing the cell to expand perpendicular to the field. Cells in (c) are migrating towards the electric field.}
	\label{fig:fourier}
\end{figure*}

Keratocytes migrate with their long axis perpendicular to the electric field. This could be because they sense more accurately along the short axis, which our circular variance results may suggest, and earlier work has also assumed \cite{kaiyrbekov2024does}. Another possibility is that the keratocyte orientation to the field is a downstream effect of the cell creating protrusions towards the electric field and then elongating perpendicular to the field. Both scenarios hint at a link between sensor redistribution and cell shape and orientation. Up to this point, we have only measured the cell's response to an electric field while considering its fixed elliptical shape. Given that cells may change shape in response to electric fields \cite{luther1983changes,onuma1988electric}, we want to link sensing mechanics with cell deformation and find out whether cells tend to expand parallel to the field or perpendicular to the field. 

We build an initial simple model of cell shape coupled with field sensing by describing the cell shape as arising from radial protrusions that are graded around the cell boundary \cite{lee1993principles}. As we discussed in the vector sum section, our idea is that the cell creates small protrusions perpendicular to its outer edge in areas where there is a high concentration of nearby sensors. As a result, cell direction and shape are controlled by protrusions normal to the boundary \cite{lee1993principles,keren2008mechanism}. We will assume (to make a few later calculations simpler) that the cell is near-circular. We can then describe the cell boundary by the radius function $\mathcal{R}(\theta,t)$ in polar coordinates (Fig. \ref{fig:fourier}a) \cite{ohta2009deformable,ohta2009deformationreaction,hiraiwa2010dynamics}, which can be expanded by Fourier series
\begin{equation}
	\mathcal{R}(\theta,t)=\mathcal{R}_0+\delta\mathcal{R}(\theta,t)=\mathcal{R}_0+\sum_{-\infty}^{\infty}\rho_n(t)e^{in\theta}.
\end{equation}
The $n=0,\pm1$ terms are implicitly excluded from the sum.  The $n=0$ mode corresponds to a uniform expansion/contraction of the cell, and the $n=\pm1$ modes correspond to simple translational motion, which we include by assuming that the cell is initially traveling with a constant speed. We assume the force balance for the cell interface can be given by 
\begin{equation}
	\tau\mathcal{V}(\theta,t)=\eta\mathcal{K}(\theta,t)+\alpha \left(c(\theta,t) - c^*\right),
\end{equation}
where $\mathcal{V}$ is the normal velocity of the cell interface, $\mathcal{K}$ is the interface's curvature, and $c$ is the local concentration of sensors. $\tau$ is a friction coefficient, i.e. $-\tau \mathcal{V}$ is the frictional drag force per unit length. The term proportional to the interface curvature tends to minimize the cell perimeter (i.e. it reflects a line tension) -- the coefficient $\eta$ has units of force. The $\alpha \left(c(\theta,t) - c^*\right)$ term says that, if $\alpha > 0$, the cell generates normally outward protrusions where the sensor concentration is above a threshold concentration, $c > c^*$, and contractions elsewhere. Together these assumptions are essentially those of the mechanical models in, e.g. \cite{shao2010computational,camley2017crawling,camley2014polarity}. Similar equations of motion can be derived for interfaces in reaction-diffusion dynamics \cite{ohta1989higher}. %

We are primarily interested in whether the shape expands parallel to or perpendicular to the field -- this is information that is encoded in the $n = \pm 2$ Fourier modes. Following \cite{ohta2009deformationreaction} and expanding the normal velocity $\mathcal{V}$ and the curvature $\mathcal{K}$ in terms of the Fourier modes, we can find an equation of motion for the $n=\pm2$ modes, 
\begin{equation}
	\label{eq:fouriermodes}
	\dfrac{\mathrm{d}\rho_{\pm2}}{\mathrm{d}t}=-\bar{\eta}\rho_{\pm2}+\dfrac{\bar{\alpha}}{2\pi}\int_{\psi-\pi}^{\psi+\pi}\mathrm{d}\theta c(\theta,t)e^{\mp i2\theta},
\end{equation}
where $\bar{\eta}=3\eta\mathcal{R}_0^{-2}\tau^{-1}$ and $\bar\alpha=\alpha\tau^{-1}$. Essentially, this equation reflects a balance of the line tension trying to make the cell more isotropic ($\bar{\eta}$ term, which decreases the amplitude of this Fourier mode) and the distribution of protrusion around the cell boundary in the $\bar{\alpha}$ term, which could make the cell more anisotropic. %
We are primarily interested in how the cell tends to break symmetry, whether it elongates parallel to or perpendicular to the field, so we will start by assuming that the cell is near-circular and so we use the formula for $c(\theta,t)$ appropriate for a circle. 
In this limit, Eq. \eqref{eq:concentration} becomes $c(\theta) = c_0 e^{\kappa \cos (\theta - \psi)}$. We will choose the normalization constant $c_0$ so that the concentration integrates to $1$, i.e. 
	$c(\theta,t)=\dfrac{e^{\kappa(t)\cos(\theta-\psi)}}{2\pi I_0(\kappa(t))}$. (This absorbs the total number of sensors into the prefactor $\alpha$.)
Calculating the integral from Eq. \eqref{eq:fouriermodes} gives us
\begin{equation}
	\dfrac{\mathrm{d}\rho_{\pm2}}{\mathrm{d}t}=-\bar{\eta}\rho_{\pm2}+\dfrac{\bar\alpha}{2\pi}\dfrac{I_2(\kappa)}{I_0(\kappa)}e^{\mp i2\psi}.
\end{equation}
We then predict that at steady state, the Fourier modes settle to
\begin{equation}
	\rho_{\pm2}=\dfrac{\bar a}{2\pi}\dfrac{I_2(\kappa)}{I_0(\kappa)}e^{\mp i2\psi},
\end{equation}
where $\bar a=\bar\alpha/\bar{\eta}$. Moving back to real space, if we only include the $n = \pm 2$ Fourier modes, we then see how the cell's radius deviates from $R_0$ as
\begin{align}
	\delta\mathcal{R}(\theta)&=\rho_2e^{i2\theta}+\rho_{-2}e^{-i2\theta}\\
 &=\dfrac{\bar a}{2}\mathfrak{r}(\kappa)\cos[2(\theta-\psi)], \label{eq:change_in_R}
\end{align} 
where $\mathfrak{r}(\kappa)=\dfrac{2}{\pi}\dfrac{I_2(\kappa)}{I_0(\kappa)}$. 
Eq. \eqref{eq:change_in_R} shows that, if $\alpha > 0$ ($\bar a>0$), the cell will expand symmetrically along the field direction (Fig. \ref{fig:fourier}c). Changing the electric field strength (changing $\kappa$) will change the magnitude of this expansion, but not its sign, as $\mathfrak{r}(\kappa)$ does not change sign when $\kappa$ varies (Fig. \ref{fig:fourier}b). 
We have so far assumed that $\alpha > 0$ implicitly by considering the mobility $\mathfrak{m}$ as nonnegative. This is a remnant of our earlier work where we assumed that sensors led to a net outward protrusion \cite{nwogbaga2023physical}. If the sensors are transported in the direction of the electric field, as we've illustrated in Fig. \ref{fig:schematicfield}, net forces in the direction of the sensors will push the cell in the direction of the field. However, there is no constraint that our mobility $\mathfrak{m}$ (and subsequently $\kappa$) must be positive \cite{mclaughlin1981role}! If sensors are instead transported to the {\it back} of the cell ($\mathfrak{m} < 0$), all of our earlier results will still hold -- the results on Fisher information, etc. are insensitive to the sign of $\kappa$. However, if the sensors are instead at the back of the cell, the ``vector sum'' must have the opposite sign -- the cell needs to generate contractile force where the sensors are highly concentrated. Therefore, for sensors that are swept to the back of the cell, we should have a negative value of $\alpha$, and thus, cells will expand perpendicular to the field (Fig. \ref{fig:fourier}c). This suggests that one possible interpretation of the observation that many cells tend to galvanotax perpendicular to the field is that they may have sensors that are swept to the cell rear. 

We note that there are many caveats on our analysis here. The first is that cell shape is not just influenced by this radial protrusion, but also by any pre-existing cell polarity. Keratocytes, for instance, are elongated whether or not there is a field applied, and this elongated shape may arise from a wide variety of different models, independent of the field \cite{keren2008mechanism,wolgemuth2011redundant,shao2010computational}. Our calculation essentially describes whether or not we would expect a net tendency for the cell to elongate perpendicular to the field. A second caveat of this calculation is that, to get an analytically tractable answer, we have restricted ourselves to assuming $c(\theta)$ is given by the steady-state limit for a circular cell. A more rigorous approach would be to initialize the cell with a uniform concentration of sensors, apply a field, and solve for the coupled transport of the sensors, shape change of the cell, and the change in the external field due to cell shape change in tandem. This is a much more complicated problem, and not within the scope of the current work. 

\section*{\label{sec:discussion}Discussion}

Are cells better sensors of an electric field with long axis parallel to the field or perpendicular to the field? We find a fairly complicated answer. Figs. \ref{fig:keratocyte} and \ref{fig:receptorsum} show that the best choice depends both on the strength of the electric field, with the choice switching within the typical field range of 10-1000 mV/mm, and whether a cell estimates the field direction with a maximum likelihood estimation (likely difficult to compute) or a vector sum (plausible to compute). %

Many groups have studied the fundamental physical limits of measurement accuracy of chemotaxis and chemosensing  \cite{hu2011geometry,ipina2022collective,fancher2017fundamental,bialek2005physical,ten2016fundamental,camley2018collective,tweedy2013distinct,nakamura2024gradient}. Often, this work implicitly assumes that the cell can compute a maximum likelihood estimate -- an estimate that may be physically intractable for the cell to compute due to its complexity. Our related work on galvanotaxis shows that simple, physically-plausible estimators like the vector sum may be biased for cells that don't have circular symmetry. This bias, however, is limited if cells' long axes are perpendicular or parallel to the field.

Many cell types migrate in an electric field with their long axis perpendicular to the field, including keratocytes, fibroblasts, multipotent mesenchymal stem cells, endothelial progenitor cells, neurons, neural crest cells, etc. \cite{mehta2021physiological,guo2015calcium,allen2013electrophoresis,allen2020cell,bai2004dc,zhao2012directing,bunn2019dc,bai2004dc,cooper1984perpendicular,rajnicek1992electric,zhao2012directing,tandon2009alignment}, though {\it Dictyostelium discoideum} is a notable exception \cite{sato2007input}. Do cells orient this way because it gives them a benefit during galvanotaxis \cite{kaiyrbekov2024does}? This is plausible -- there may be large increases in accuracy if cells are correctly oriented, especially if they are using the vector sum estimator (Fig. \ref{fig:receptorsum}d). However, cells may also just naturally travel with their long axis parallel to their direction of polarity even in the absence of field -- as keratocytes do. The most dramatic example of cell reorientation and shape change is Schwann cells, which would normally migrate with the long axis parallel to their direction of travel in the absence of a field, switch orientation to have the long axis perpendicular when in the presence of an electric field \cite{lang2021use}.

A second possibility explaining why cells tend to migrate with long axes perpendicular to the field is that this orientation occurs as a side effect of a mechanism where sensors drive protrusion. Our results show that a cell would expand parallel to the field if the sensors are in the front of the cell $(\kappa>0)$, and expand perpendicular to the field if the sensors are at the back of the cell $(\kappa<0)$. If this simple mechanism is reasonable, cell shape would directly indicate whether sensors localize to the cell front or cell back. Because Schwann cells \cite{lang2021use} and MDCK cells \cite{shim2021overriding} expand perpendicular to an applied field but do not have this orientation in the absence of field, our model would predict that they have sensors that localize to the cell back. %

Our results in Fig. \ref{fig:circularvariance} show something somewhat disturbing and interesting for theorists of fundamental limits: the maximum likelihood estimator has a circular variance that is not only larger than our modified Cramer-Rao bound, but which has an {\it opposite} trend to the bound at small field strengths. While it is well known that the maximum likelihood estimator is only efficient (reaching the Cramer-Rao limit) in the limit of large numbers of samples \cite{kay1993fundamentals} (here, large numbers of sensors), it is striking that the Fisher information can be completely misleading. Within our calculations, the Fisher information is always maximal when the electric field is oriented parallel to the cell's long axis. However, the maximum likelihood estimator's variance can be maximal in this circumstance -- in other words, the variance of the maximum likelihood estimator is not even a monotonic function of the Fisher information. This result may encourage some revisiting of earlier works applying maximum likelihood estimation and related Cramer-Rao bounds \cite{endres2009maximum,hopkins2020chemotaxis,hu2011geometry,ipina2022collective}, and more emphasis on understanding the details of how cells can compute these estimates \cite{mehta2012energetic,singh2017simple} -- which are not well understood for gradient sensing. In particular, earlier results on chemotaxis in elliptical cells \cite{hu2011geometry} may potentially have similar discrepancies between MLE and Cramer-Rao. The Monte Carlo simulations within \cite{hu2011geometry} do show agreement -- but it is not clear whether this would hold in the more experimentally relevant regime where the difference between the circular variance and variance matters. %

\begin{acknowledgments}
{We acknowledge support from NSF PHY 1915491 and NIH R35 GM142847. %
We thank Emiliano Perez Ipi\~na and Daiyue Sun for a close reading of the draft.}
\end{acknowledgments}

\renewcommand{\thefigure}{S\arabic{figure}}
\setcounter{figure}{0}

\onecolumngrid
\appendix

\section*{Supplementary Information: ``Cell shape and orientation control galvanotactic accuracy''}

\section{Elliptic coordinates fundamentals}
\label{app:ellipticcoordinates}

\begin{figure}[htb]
	\centering
	\includegraphics[width=0.6\linewidth]{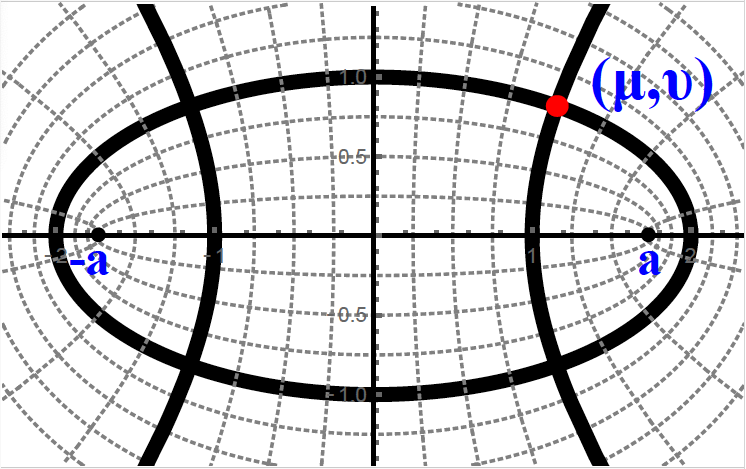}
	\caption{Elliptic coordinate system displaying confocal ellipses and hyperbolae. Every point corresponds to the intersection of an ellipse ($\mu$) and a hyperbola branch ($\nu$). The foci for the ellipses and hyperbolae are at $\pm a$.}
	\label{fig:ellipticcoordinates}
\end{figure}

Elliptic coordinates are a standard orthogonal coordinate system that generalizes polar coordinates \cite{hobson1931theory,korn2000mathematical,morse1954methods,darboux1896leccons}. In elliptic coordinates every point is determined by an elliptic radius $\mu$ and elliptic angle $\nu$, analogous to the radius and polar angle in polar coordinates. A constant elliptic radius lies on the boundary of an ellipse, while a constant elliptic angle lies on a hyperbola that is horizontally oriented (Fig. \ref{fig:ellipticcoordinates}). Ellipses and hyperbolae centered at the origin can be parameterized by $(R_1\cos t,R_2\sin t)$ and $(\alpha_1\sec t,\alpha_2\tan t)$, respectively. $R_1, R_2$ are the semi-major and minor axes of the ellipse. $2\alpha_1$ is the distance between the two vertices of the hyperbola. $\alpha_2^2=a^2-\alpha_1^2$, where $a$ represents the focus point of the ellipses and hyperbolae, $a=\sqrt{R_1-R_2}$. $t$ is the variable of parameterization, where $t\in[0,2\pi]$. The angle between the x-axis and the hyperbola asymptote is $\nu$, which can be computed from the equation $\nu=\arctan{[2\alpha_1\alpha_2/(\alpha_1^2-\alpha_2^2)]}$.

There are multiple conventions for representing Cartesian variables with elliptic variables. In this manuscript, we use the convention where $x=a\cosh\mu\cos\nu$ and $y=a\sinh\mu\sin\nu$. The unit vectors and scale factors are 
\begin{equation}
	\bm{\hat{\mu}}=\dfrac{a}{h_{\mu}}\begin{pmatrix}
		\sinh\mu\cos\nu\\
		\cosh\mu\sin\nu
	\end{pmatrix},\quad
	\bm{\hat{\nu}}=\dfrac{a}{h_{\nu}}\begin{pmatrix}
		-\cosh\mu\sin\nu\\
		\sinh\mu\cos\nu
	\end{pmatrix},\quad h_{\mu}=h_{\nu}=a\sqrt{\dfrac{\cosh{2\mu}-\cos{2\nu}}{2}}.
\end{equation}
Note $h_{\mu}$ and $h_{\nu}$ are equal. For our elliptical cell, placing the cell boundary at elliptical parameter $\mu=\mu_0$ means that the semi-major and semi-minor axes are $R_1=a\cosh\mu_0$ and $R_2=a\sinh\mu_0$. $\lambda = R_1/R_2$ is the aspect ratio. 

To produce the field lines in Fig. \ref{fig:fieldlines} in the main text, it is sometimes easier to produce a grid in polar coordinates first, then to obtain expressions for both $\mu$ and $\nu$ in terms of the polar coordinates $r$ and $\theta$. There are established relationships we can use to map between Cartesian, polar, and elliptic coordinates, which we summarize here. We start with $x=r\cos\theta=a\cosh\mu\cos\nu$ and $y=r\sin\theta=a\sinh\mu\sin\nu$. Using Euler's formula and complex trigonometric relations $\cos\nu=\cosh{i\nu}$ and $i\sin\nu=\sinh{i\nu}$, we can derive a relationship between polar and elliptic coordinates.
\begin{equation}
	\begin{split}
		re^{i\theta}&=r\left(\cos\theta+i\sin\theta\right)=a\left(\cosh\mu\cos\nu+i\sinh\mu\sin\nu\right)\\
		&=a\left(\cosh\mu\cosh{i\nu}+\sinh\mu\sinh{i\nu}\right)=a\cosh{(\mu+i\nu)}.
	\end{split}
\end{equation}
We now see that there is a straightforward relationship between $(r,\theta)$ and $(\mu,\nu)$:
\begin{align}
	\mu(r,\theta)&=\Re\left[\arccosh{\left(\dfrac{r}{a}e^{i\theta}\right)}\right],\\[5pt]
	\nu(r,\theta)&=\Im\left[\arccosh{\left(\dfrac{r}{a}e^{i\theta}\right)}\right],
\end{align}
where $\Re(\cdot)$ and $\Im(\cdot)$ are the real and imaginary parts, respectively. The solutions are \cite{dassios1989capacity}
\begin{align}
	\mu&=\arccosh{\left(\dfrac{\mathfrak{A}_++\mathfrak{A}_-}{2a}\right)},\\
	\nu&=\dfrac{\sin\theta}{\vert\sin\theta\vert}\arccos{\left(\dfrac{\mathfrak{A}_+-\mathfrak{A}_-}{2a}\right)},
\end{align}
where $\mathfrak{A}_\pm=\sqrt{r^2\pm2ra\cdot\cos\theta+a^2}$. An equivalent result for $\nu$ has been derived \cite{kotnik2000analytical}
\begin{equation}
	\nu=\dfrac{\sin\theta}{\vert\sin\theta\vert}\arccos{\left(\dfrac{r}{R_1}\cos\theta\right)}=\dfrac{\sin\theta}{\vert\sin\theta\vert}\arccos{\left(\dfrac{R_1R_2\cos\theta}{R_1\sqrt{(R_1\sin\theta)^2+(R_2\cos\theta)^2}}\right)}=\dfrac{\sin\theta}{\vert\sin\theta\vert}\arccos{\left(\dfrac{\cos\theta}{\sqrt{\lambda^2\sin^2\theta+\cos^2\theta}}\right)},
\end{equation}
where $\lambda=R_1/R_2$ is defined as the ratio between the semi-major axis and the semi-minor axis lengths.

 For completeness, we can also map from elliptic coordinates back to polar coordinates. We know that $r^2=x^2+y^2$ and $y/x=\tan\theta$. If we make use of the identities $\cos^2\nu+\sin^2\nu=1$ and $\cosh^2\mu-\sinh^2\mu=1$, we arrive at the identities \cite{morse1954methods}
\begin{align}
	r & = a\sqrt{\left(\cosh^2{\mu}-\sin^2{\nu}\right)},\\
	\theta &= \arctan{(\tanh\mu\tan\nu)} .
\end{align}

\section{Boundary conditions on the electric field}
\label{app:boundary}

We assume a no-flux-like boundary condition, i.e. that there is no electric field normal to the membrane of the cell, $\mathbf{E}\cdot\hat{\mathbf{n}} = 0$. This boundary condition is physically motivated by the high resistivity (poor conductivity) of the cell membrane. The resistivity of the cell membrane is high (10$^6$--10$^8$ $\Omega\cdot$m)  \cite{kotnik2000analytical,kotnik1997sensitivity,cartee1992transient,kotnik2000second} compared to the resistivity of the surrounding aqueous solution ($\sim$0.008--0.8 $\Omega\cdot$m \cite{debruin1999modeling,kotnik2000second}) and the cytoplasm (3 $\Omega\cdot$m \cite{kotnik2000second}). Why does this lead to a no-flux boundary condition? One form of Gauss's law is
\begin{equation}
	\nabla\cdot\mathbf{D}=\rho_f,
\end{equation}
where $\mathbf{D}$ is the displacement field and $\rho_f$ is the free charge density. In linear media, $\mathbf{D}=\epsilon\mathbf{E}$. Using this relation and the definition $\mathbf{E}=-\nabla\Phi$, we can recast Gauss' law:
\begin{equation}
	-\nabla\cdot\left(\epsilon\nabla\Phi\right)=\rho_f.
\end{equation}
Taking the time derivative on both sides and using the continuity equation $\partial_t\rho_f+\nabla\cdot\mathbf{J}_f=0$, we can again rewrite Gauss' law:
\begin{equation}
	-\nabla\cdot\left(\epsilon\nabla\dfrac{\partial\Phi}{\partial t}\right)=-\nabla\cdot\mathbf{J}_f,
\end{equation}
where $\mathbf{J}_f$ is the current density. At steady state, $\nabla\cdot\mathbf{J}_f=0$. Ohm's law tells us that $\mathbf{J}_f=\sigma\mathbf{E}=-\sigma\nabla\Phi$. $\sigma$ is the conductivity. This means that for Ohmic materials, the potential will obey \cite{pucihar2009time}
\begin{equation}
	0=\nabla\cdot\left(\epsilon\nabla\dfrac{\partial\Phi}{\partial t}\right)+ \nabla\cdot \left(\sigma \nabla \Phi\right).
\end{equation}
We assume that there is an effective steady state, the electric field is not changing rapidly, so we neglect the first term. Thus, the potential -- inside the cell, outside the cell, or on the membrane -- will obey
\begin{equation}
\nabla\cdot \left(\sigma \nabla \Phi\right)=0.
\end{equation}
We can use this to establish the boundary conditions. To be in steady state, the current flux across a boundary must be continuous, so at the membrane-exterior fluid boundary, the field must obey
\begin{align}
\mathbf{J}^{\textrm{membrane}}_f\cdot\mathbf{\hat{n}}&=\mathbf{J}^{\textrm{fluid}}_f\cdot\mathbf{\hat{n}},\\	\sigma_\textrm{membrane}\left(\nabla\Phi\right)\cdot\mathbf{\hat{n}}&=\sigma_\textrm{fluid}\left(\nabla\Phi\right)\cdot\mathbf{\hat{n}},
\end{align}
If the conductivity of the membrane can be neglected, $\sigma_\textrm{membrane} \approx 0$, then the boundary condition at the membrane surface reduces to simply $\nabla\Phi\cdot\mathbf{\hat{n}} = 0$.  %

\section{Electric field around a circular cell}
\label{app:electricfieldcircle}

Before finding the electric field for the elliptical cell, doing so for a circular cell will provide some intuition. We solve Laplace's equation in cylindrical coordinates. We will assume that the potential is zero along the axial direction. Under this asumption, Laplace's equation takes the form:
\begin{equation}
	\nabla^2\Phi=\dfrac{1}{r}\dfrac{\partial}{\partial r}\left(r\dfrac{\partial\Phi}{\partial r}\right)+\dfrac{1}{r^2}\dfrac{\partial^2\Phi}{\partial\theta^2}=0.
\end{equation}
The most general form of the potential is \cite{griffiths2023introduction}  
\begin{equation}
\Phi(r,\theta)=a_0+a_1\ln{r}+\sum_{k=1}^{\infty}\left(A_kr^k+B_kr^{-k}\right)\left(C_k\cos k\theta+D_k\sin k\theta\right).
\end{equation}
We can choose a reference point for our potential, so we can set $a_0=0$. Far away from the cell, we know that $\Phi=-E_0(x\cos\psi+y\sin\psi)\equiv-E_0r(\cos\theta\cos\psi+\sin\theta\sin\psi)=-E_0r\cos(\theta-\psi)$. This implies that the $r^1$ term is nonzero, while the remaining $r^k$ terms go to zero. The $r^{-k}$ terms are permitted, since we are only solving for the potential outside the cell, so the divergence at the origin is not a problem. 
\begin{equation}
	\Phi(r,\theta)=-E_0r(\cos\theta\cos\psi+\sin\theta\sin\psi)+\sum_{k=1}^{\infty}\left(C_k\cos k\theta+D_k\sin k\theta\right)B_kr^{-k},
\end{equation}
subject to the boundary condition $\partial_r\Phi=0$ when $r=R_0$. 

Applying the boundary condition to the potential on the outside of the cell gives us:
\begin{equation}
	-E_0(\cos\theta\cos\psi+\sin\theta\sin\psi)-\sum_{k=1}^{\infty}k\left(C_k\cos k\theta+D_k\sin k\theta\right)B_kR_0^{-k-1}=0.
\end{equation}
Matching the trigonometric terms order-by-order, all $k\neq1$ terms are zero, and we find
$-E_0\cos\psi=C_1B_1R_0^{-2}$ and $-E_0\sin\psi=D_1B_1R_0^{-2}$.
We can then use this to rewrite the potential $\Phi$ as
\begin{align}
	\Phi(r,\theta)&=-E_0r\cos(\theta-\psi)-E_0\dfrac{R_0^2}{r}\cos(\theta-\psi).
\end{align}
The electric field can be found as $\mathbf{E}=-\nabla\Phi$:
\begin{align}
\mathbf{E}&=E_0\left(1-\dfrac{R_0^2}{r^2}\right)\cos(\theta-\psi)\mathbf{\hat{r}}-E_0\left(1+\dfrac{R_0^2}{r^2}\right)\sin(\theta-\psi)\bm{\hat{\theta}}.
\end{align}
We see that, as we demanded, far from the cell the field is simply the applied electric field $\mathbf{E}_\mathrm{ext}=E_0\mathbf{\hat{E}}$. However, close to the cell, the field deforms around the cell boundary -- choosing $r = R_0$ we see that the field is completely tangential at the cell's surface, as required by our boundary condition. 
It is easy to see that the field tangent to the cell at the cell surface $r=R_0$ is
\begin{equation}
	\mathbf{E}_\parallel=-2E_0\sin(\theta-\psi)\bm{\hat{\theta}}.
\end{equation}

\section{Electric field around an elliptical cell}
\label{app:electricfieldellipse}

We want to calculate the electric field along the boundary of the cell. To accomplish this, we solve for the tangential electric field in elliptical cylindrical coordinates. That first requires solving Laplace's equation $\nabla^2\Phi=0$ in 2D. Laplace's equation in elliptical coordinates is:
\begin{equation}
	\nabla^2\Phi=\dfrac{1}{h_\mu h_\nu}\left(\dfrac{\partial^2\Phi}{\partial\mu^2}+\dfrac{\partial^2\Phi}{\partial\nu^2}\right)=0.
\end{equation}
$h_\mu=h_\nu$ are scale factors and $a=\sqrt{R_1^2-R_2^2}$, assuming $R_1>R_2$, where $R_1$ and $R_2$ are the lengths of the semi-major and minor axes, respectively. We can multiply both sides of Laplace's equation by $h_\mu h_\nu$ to simplify further. Let $\Phi(\mu,\nu)=M(\mu)N(\nu)$. By substituting and dividing through by $\Phi$, we get:
\begin{equation}
	\dfrac{1}{M}\dfrac{\partial^2M}{\partial\mu^2}=-\dfrac{1}{N}\dfrac{\partial^2N}{\partial\nu^2}=k^2.
\end{equation}
By inspection, solutions for $N$ have the form $C_ke^{ik\nu}$. For $M$, we have to consider two cases. If $k=0$, $M(\mu)=c_1\mu+c_0$. If $k\neq0$, it is advantageous to guess a solution in the form $M(\mu)=c_{2k}\cosh{k\mu}+c_{3k}\sinh{k\mu}+c_{4k}e^{k\mu}+c_{5k}e^{-k\mu}$. 
This produces a general solution of the form
\begin{equation}
	\Phi(\mu,\nu)=c_0+c_1\mu+\sum_{k=1}^{\infty}C_ke^{ik\nu}[c_{2k}\cosh{k\mu}+c_{3k}\sinh{k\mu}+c_{4k}e^{k\mu}+c_{5k}e^{-k\mu}].
\end{equation}
Just like for a circular cell, the cell is in a uniform applied field. We know that, far away from the cell $(\mu\rightarrow\infty)$, the electric field potential outside should take the form $\Phi=-E_0(x\cos\psi+y\sin\psi)=-aE_0(\cosh\mu\cos\nu\cos\psi+\sinh\mu\sin\nu\sin\psi)$. This implies that our solution should take the form
\begin{equation}
	\Phi=-aE_0(\cosh\mu\cos\nu\cos\psi+\sinh\mu\sin\nu\sin\psi)+\sum_{k=1}^{\infty}\left(\mathfrak{c}_k\cos{k\nu}+\mathfrak{d}_k\sin{k\nu}\right)e^{-k\mu}.
\end{equation}
Similar to the circle case, we have the boundary condition 
\begin{align}
	\dfrac{\partial\Phi}{\partial\mu}\bigg\vert_{\mu=\mu_0}&=0.
\end{align}
We can see by matching trigonometric terms order-by-order that all terms with $k\neq1$ are zero. The boundary condition then tells us that
\begin{align}
	-aE_0(\sinh\mu_0\cos\nu\cos\psi+\cosh\mu_0\sin\nu\sin\psi)=\left(\mathfrak{c}_1\cos\nu+\mathfrak{d}_1\sin\nu\right)e^{-\mu_0}. 
\end{align}
Collecting like terms for $\cos\nu$ and $\sin\nu$,%
\begin{align}
	\mathfrak{c}_1=-aE_0e^{\mu_0}\sinh\mu_0\cos\psi,\\[10pt]
	\mathfrak{d}_1=-aE_0e^{\mu_0}\cosh\mu_0\sin\psi.
\end{align}
With substitution and some algebraic manipulations, the potential is expressed as:
\begin{align}
	\Phi&=-E_0a\left(\cosh\mu+\sinh\mu_0e^{\mu_0-\mu}\right)\cos\nu\cos\psi-E_0a\left(\sinh\mu+\cosh\mu_0e^{\mu_0-\mu}\right)\sin\nu\sin\psi.
\end{align}
(This is a well-known result; see Eq. 10.1.28 of %
\cite{morse1954methods}.)

\begin{figure}[htb]
	\centering
	\includegraphics[width=0.6\linewidth]{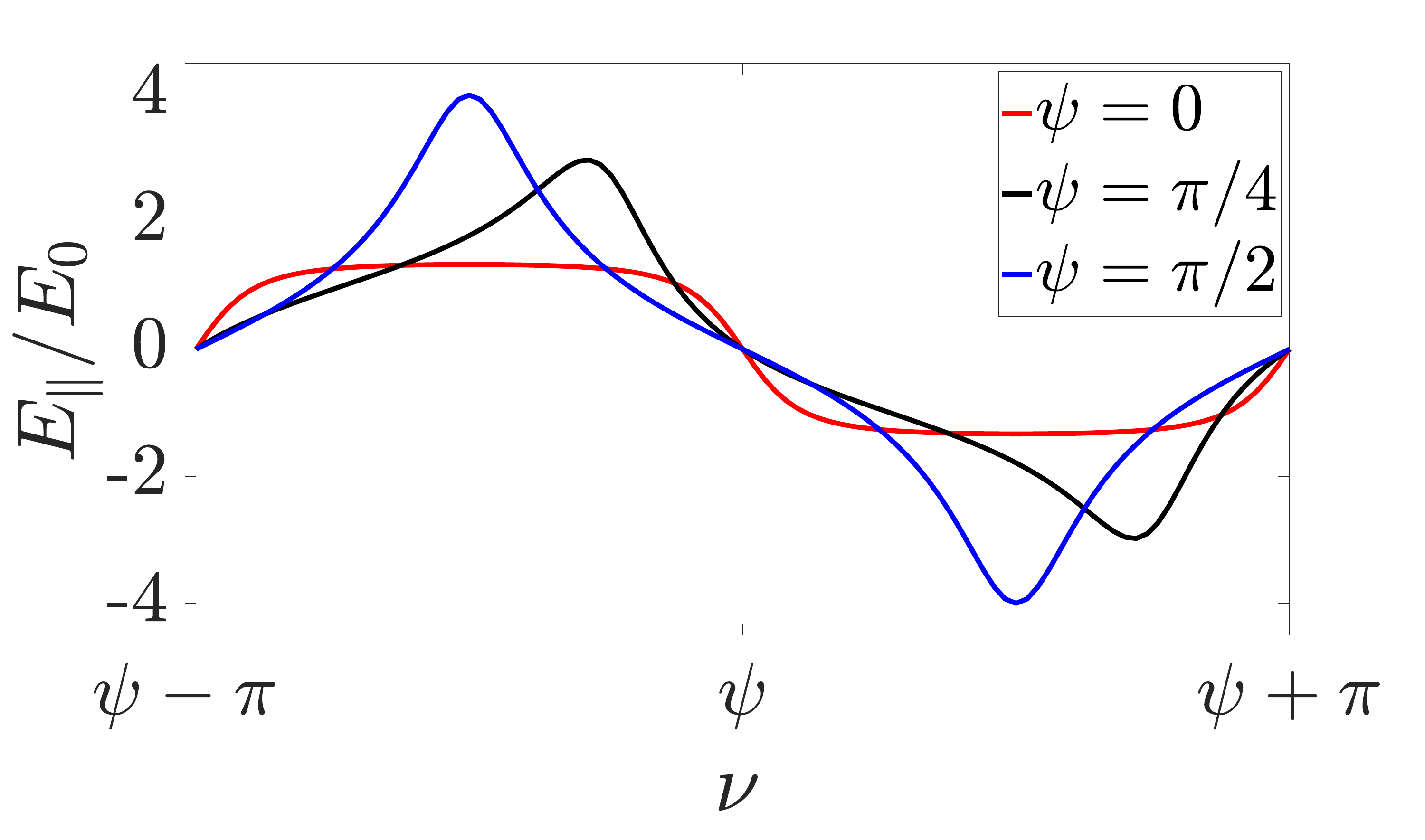}
	\caption{Plot of magnitude of $\mathbf{E}_\parallel$ normalized by $E_0$. Discussed in the main text, this is proportional to the speed of sensors on the cell surface. A positive value means the tangential field is pointing along $+\bm{\hat\nu}$, which is defined in the counterclockwise direction.}
	\label{fig:tangentialfield}
\end{figure}

The electric field outside the cell can be easily derived by taking the gradient in elliptic coordinates:
\begin{equation}
		\mathbf{E}=-\nabla\Phi=-\dfrac{1}{h_{\mu}}\dfrac{\partial\Phi}{\partial\mu}\bm{\hat{\mu}}-\dfrac{1}{h_{\nu}}\dfrac{\partial\Phi}{\partial\nu}\bm{\hat{\nu}}.
\end{equation}
The outer field in its full form is
\begin{equation}
	\mathbf{E}=\dfrac{E_0a}{h_\mu}\dfrac{e^\mu-e^{2\mu_0-\mu}}{2}\cos(\nu-\psi)\bm{\hat{\mu}}-\dfrac{E_0a}{h_\mu}\dfrac{e^\mu+e^{2\mu_0-\mu}}{2}\sin(\nu-\psi)\bm{\hat{\nu}}.
\end{equation}
It is easy to see that the field tangent to the cell at the cell surface $\mu=\mu_0$ is
\begin{equation}
	\mathbf{E}_\parallel=-E_0\dfrac{ae^{\mu_0}}{h_{\mu_0}}\sin(\nu-\psi)\bm{\hat{\nu}},
\end{equation}
where $h_{\mu_0}$ is $h_{\mu}$ evaluated at $\mu=\mu_0$. Fig. \ref{fig:tangentialfield} plots Eq.  \eqref{eq:tangential} for different field directions $\psi$. 

As a quick sanity check, does the result reduce to what we found for a circle? Note that for a circle, $a\rightarrow0$ as $\mu_0\rightarrow\infty$. That manifests as $a\cosh\mu_0=a\sinh\mu_0=R_0$, $h_{\mu_0}=a\sqrt{(\cosh{2\mu_0}-\cos{2\nu})/2}=R_0$, and $ae^{\mu_0}=a(\cosh\mu_0+\sinh\mu_0)=2R_0$. In this limit, the cell boundary is $x = R_0\cos \nu$ and $y = R_0\sin\nu$, so the elliptic angle $\nu$ is the same as the polar angle $\theta$, and the tangential electric field is
\begin{equation}
	\mathbf{E}_\parallel=-2E_0\sin(\theta-\psi)\bm{\hat{\theta}}.
\end{equation}
as we found above. 

\section{Probability distribution}
\label{app:probdist}

\begin{figure*}[htb]
	\centering
	\includegraphics[width=\textwidth]{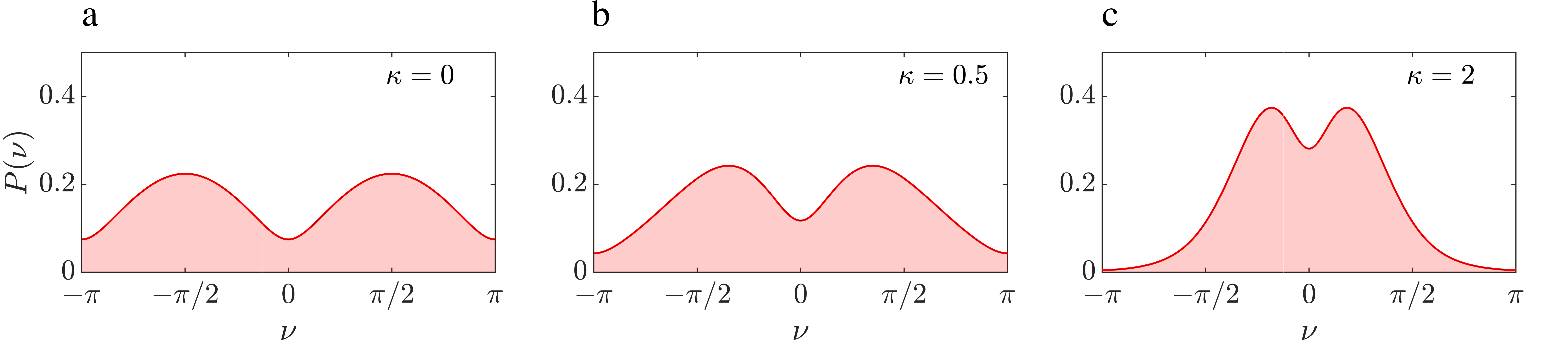}
	\caption{Probability distributions at $\psi=0$ as a function of the elliptic angle $\nu$. The cell aspect ratio $\lambda=3$. (a): $\mathcal{P}(\nu)$ for $\kappa=0$. This is a uniform distribution on an ellipse. (b): $\mathcal{P}(\nu)$ for $\kappa=0.5$. (c): $\mathcal{P}(\nu)$ for $\kappa=2$.}
	\label{fig:probdist}
\end{figure*}

\begin{figure}[htb]
	\centering
	\includegraphics[width=0.6\linewidth]{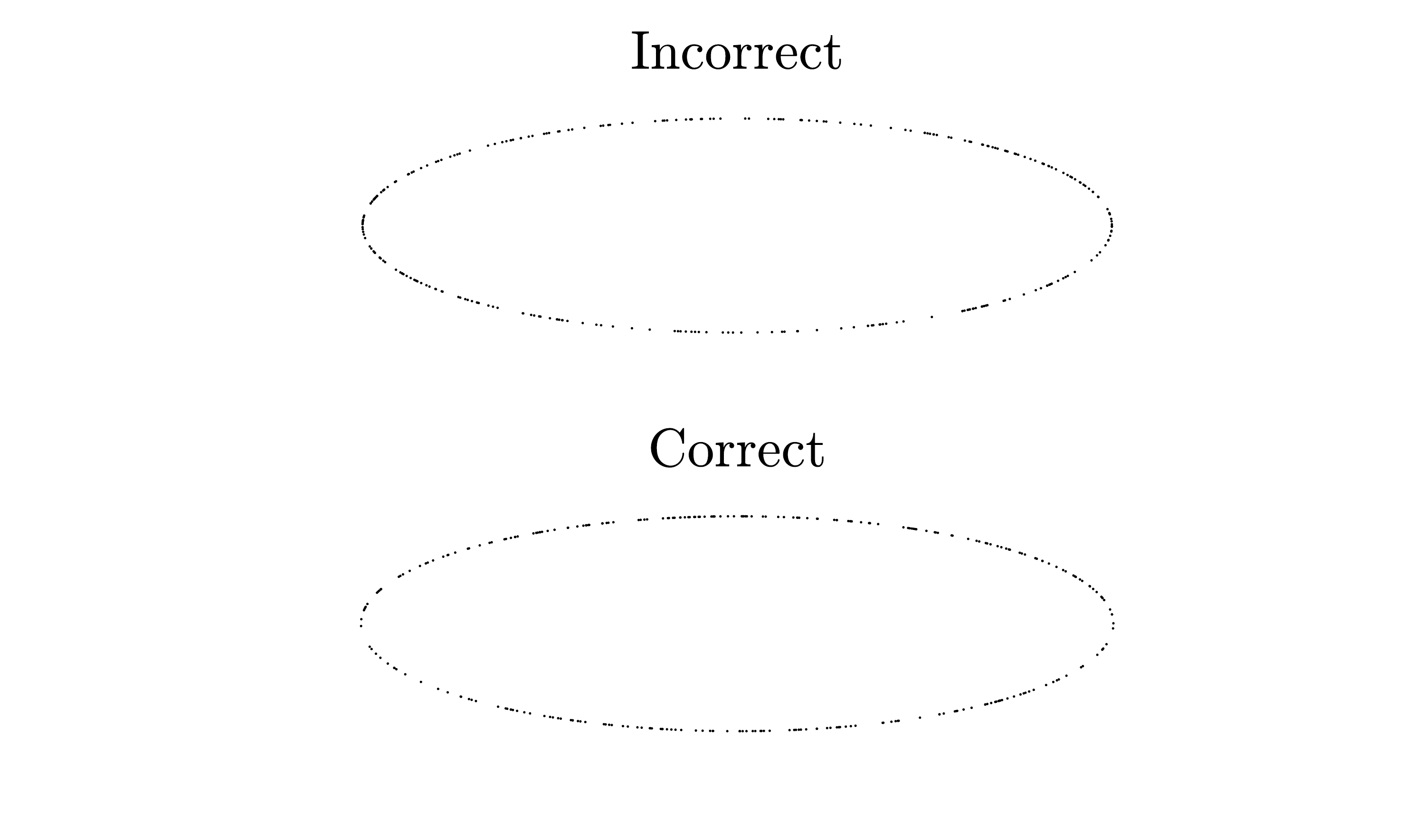}
	\caption{Showing points sampled on ellipse. Incorrect sampling leads to higher density of points along high curvature regions.}
	\label{fig:sampling}
\end{figure}

We have derived the concentration of sensors $c(\nu) = c_0 \exp \left[\kappa \cos (\nu-\psi)\right]$, the number of sensors per unit length of the membrane, in the main text. From this, we want to determine the probability density function for the elliptic angle $\nu$ of sensors. We know that the probability $\mathscr{P}(\nu)\mathrm{d}\nu$ to be in the region $\nu\cdots\nu+\mathrm{d}\nu$ is proportional to $c(\nu)\mathrm{d}l$ -- where $\mathrm{d}l$ is the amount of arclength in this region. Changing variables, we expect 
\begin{align}
    \mathscr{P}(\nu)\mathrm{d}\nu &\sim c(\nu)\mathrm{d}l  \\
    &= c(\nu) \left\vert\dfrac{\mathrm{d}l}{\mathrm{d}\nu}\right\vert\mathrm{d}\nu \\[5pt]
    &= c(\nu) h_{\mu_0}\mathrm{d}\nu.
\end{align}
We want our probability density to be normalized  $\int_\gamma\mathscr{P}(\nu)\mathrm{d}\nu=1$, where the integral is evaluated over the region $[\psi-\pi,\psi+\pi]$. %
We then get 
\begin{equation}
	\mathscr{P}(\nu)=Z^{-1}c(\nu)h_{\mu_0}, \label{eq:probdensity_appendix}
\end{equation}
where 
\begin{equation}
	Z=\int_\gamma c(\nu)h_{\mu_0}\mathrm{d}\nu.
\end{equation}
Fig. \ref{fig:probdist} shows examples of this distribution for different values of $\kappa$. We note that even the limit of a uniform distribution ($\kappa = 0$) does not appear trivial -- but this is only because of the complicated relationship between the perimeter of the cell and the elliptic angle $\nu$. Taking this into account is essential to correctly sample sensor locations. If you, instead of sampling from Eq. \eqref{eq:probdensity_appendix}, attempted to generate uniformly distributed points by sampling $\nu$ uniformly over $[0,2\pi]$, you would get an incorrect distribution of points (Fig. \ref{fig:sampling}).

\section{Maximum likelihood and Fisher information}
\label{app:MLE_FI}

In the main text, we constructed the log likelihood function 
\begin{equation}
	\label{eq:loglikelihoodgeneral}
	\ln{\mathcal{L}}=\sum_{i=1}^{N}\left[\kappa\cos{(\nu_i-\psi)}+\ln{h^{(i)}_{\mu_0}}\right]-N\ln{Z},
\end{equation}
and derived an expression for the estimator $\hat\psi$ by differentiating with respect to $\psi$. From here, we can also derive an expression for the Fisher information:
\begin{equation}
	\mathcal{I}=\left\langle-\dfrac{\partial^2\ln{\mathcal{L}}}{\partial\psi^2}\right\rangle=N\left[\dfrac{1}{Z}\dfrac{\partial^2Z}{\partial\psi^2}-\left(\dfrac{1}{Z}\dfrac{\partial Z}{\partial\psi}\right)^2\right]+N\kappa\left\langle\cos{(\nu-\psi)}\right\rangle.
\end{equation}
We can easily compute from Eq. \eqref{eq:Z}
\begin{equation}
	\left(\dfrac{1}{Z}\dfrac{\partial Z}{\partial\psi}\right)^2=\kappa^2\left\langle\sin{(\nu-\psi)}\right\rangle^2.
\end{equation}
With some work, we can also see that
\begin{equation}
	\dfrac{1}{Z}\dfrac{\partial^2Z}{\partial\psi^2}=\kappa^2\left\langle\sin^2(\nu-\psi)\right\rangle-\kappa\left\langle\cos{(\nu-\psi)}\right\rangle.
\end{equation}
By combining the results and doing some algebra, we arrive at the Fisher information introduced in the main text:
\begin{equation}
\mathcal{I}=N\kappa^2\left[\left\langle\sin^2{(\nu-\psi)}\right\rangle-\left\langle\sin{(\nu-\psi)}\right\rangle^2\right].
\end{equation}
This result simplifies in the appropriate limits. For a circle, we know that $\left\langle\sin{(\theta-\psi)}\right\rangle=0$ \cite{nwogbaga2023physical}.
\begin{equation}
\mathcal{I}=N\kappa^2\left\langle\sin^2{(\theta-\psi)}\right\rangle=\dfrac{N\kappa^2}{2\pi I_0(\kappa)}\int_\gamma\sin^2{(\theta-\psi)}\exp{[\kappa\cos(\theta-\psi)]}\mathrm{d}\theta. \; \; \textrm{(circle)}
\end{equation}
In this limit, the Fisher information becomes 
\begin{equation}
	\mathcal{I}=\dfrac{N\kappa}{I_0(\kappa)}\left[\dfrac{\kappa I_0(\kappa)-\kappa I_2(\kappa)}{2}\right]=N\kappa\dfrac{I_1(\kappa)}{I_0(\kappa)} \; \; \textrm{(circle)},
\end{equation}
which is what we expect for a circle.

\section{Perturbation calculations}
\label{app:perturbation_calculations}

To determine the maximum likelihood estimator for $\psi$ and compute the Fisher information $\mathcal{I}$, we need to evaluate integrals for $Z$ and $\partial Z/\partial \psi$, where $Z = \int_{\psi-\pi}^{\psi+\pi} c(\nu) h_{\mu_0} \textrm{d}\nu$ as given in Eq. \eqref{eq:Z}. These are then used in in Eqs. \eqref{eq:MLE} and \eqref{eq:FisherInformation}. We can evaluate these integrals numerically, and do so to compute the numerical bounds, but there is no general analytical solution. However, in this Appendix \ref{app:perturbation_calculations}, we derive simpler expressions for both the MLE and the Fisher information in certain relevant limits. In Appendix \ref{app:MLE_FI_pert}, we assume that cells are nearly circular, simplifying the MLE in Eq. \eqref{eq:MLE} and providing one simplified expression for the Fisher information in Eq. \eqref{eq:FisherInformation}. In Appendix \ref{app:FI_smallkappa}, we derive two more simplified expressions for the Fisher information, one assuming that electric fields are weak, and another incorporating both assumptions, assuming that cells are nearly circular and that fields are weak simultaneously. In Appendix \ref{app:FI_approx}, we summarize all the approximations for the Fisher information and explore some of their limiting behaviors.

\subsection{Perturbation calculations for MLE and Fisher information for nearly circular cells}
\label{app:MLE_FI_pert}

To derive a tractable analytical solution, we can assume that our cells are not too elongated. This approximation allows us to make a perturbation argument that simplifies the integrals and derive an expression for both the MLE in Eq. \eqref{eq:MLE} and the Fisher information in Eq. \eqref{eq:FisherInformation}. The scale factor $h_{\mu_0}$ contains a square root that makes the integrals intractable. (Even integrating the scale factor by itself yields elliptic integrals!). Fortunately, it can be simplified greatly with a power series expansion. First, it is advantageous to rewrite the scale factor:
\begin{equation}
	h_{\mu_0}=a\sqrt{\dfrac{\cosh2\mu_0-\cos2\nu}{2}}=\sqrt{\dfrac{a^2\cosh2\mu_0}{2}}\sqrt{1-\left(\dfrac{\cos2\nu}{\cosh2\mu_0}\right)}.
\end{equation}
If the cell is nearly circular, $\cosh2\mu_0$ gets very large as $\mu_0$ tends to infinity (required as $\lambda \to 1$), %
allowing us to Taylor expand the scale factor to first order in $\left(\cosh2\mu_0\right)^{-1}$:
\begin{equation}
	h_{\mu_0}\approx R_0\left[1-\dfrac{1}{2}\left(\dfrac{\cos2\nu}{\cosh2\mu_0}\right)-\cdot\cdot\cdot\right].
\end{equation}

We will start by evaluating the partition function $Z$ in the limit of a near-circular cell. We see
\begin{align}
Z &= \int_\gamma c(\nu) h_{\mu_0} \textrm{d}\nu \\
  &\approx  R_0\int_\gamma\exp{[\kappa\cos(\nu-\psi)]}\left[1-\dfrac{1}{2}\left(\dfrac{\cos2\nu}{\cosh2\mu_0}\right)\right]\mathrm{d}\nu.
\end{align}
The integral over $\gamma$ is a shorthand for the integral over the  $2\pi$ range of $\nu$, usually from $\psi-\pi$ to $\psi+\pi$. We also took the normalization factor $c_0 = 1$ without loss of generality -- this will drop out of any probabilities. We can integrate this using some standard results for modified Bessel functions and using a substitution to $\alpha = \nu - \psi$, finding
\begin{align}
  Z&\approx2\pi R_0\left[I_0(\kappa)-\dfrac{1}{2}\dfrac{\cos2\psi}{\cosh2\mu_0}I_2(\kappa)\right]\\
  &\equiv\chi(1-\epsilon),
\end{align}
where $\chi=2\pi R_0I_0(\kappa)$ and $\epsilon=\cos2\psi I_2(\kappa)[2I_0(\kappa)\cosh2\mu_0]^{-1}$. 

Similarly, we can evaluate the derivative $\frac{\partial Z}{\partial \psi}$, which is used in Eq. \eqref{eq:MLE}, using the approximation for $h_{\mu_0}$:
\begin{equation}
	\dfrac{\partial Z}{\partial\psi}\approx R_0\int_\gamma\kappa\sin{(\nu-\psi)}\exp{[\kappa\cos(\nu-\psi)]}\left[1-\dfrac{1}{2}\left(\dfrac{\cos2\nu}{\cosh2\mu_0}\right)\right]\mathrm{d}\nu.
\end{equation}
In this formula, the zeroth-order term in $(\cosh2\mu_0)^{-1}$ will vanish because the integrand is odd. The next term can be evaluated similarly to the earlier integral,
\begin{equation}
\dfrac{\partial Z}{\partial\psi}\approx 2\pi R_0 \frac{\sin2\psi}{\cosh2\mu_0} I_2(\kappa).
\end{equation}

To finish the calculation of $\frac{1}{Z} \frac{\partial Z}{\partial \psi}$, which is on the right hand side of Eq. \eqref{eq:MLE} for nearly circular cells, we expand $Z^{-1}$ in the limit of nearly circular cells (which corresponds to small $\epsilon$):
\begin{equation}
	\dfrac{1}{Z}\approx\dfrac{1}{\chi(1-\epsilon)}\approx\dfrac{1}{\chi}\left(1+\epsilon+O(\epsilon^2)\right).
\end{equation}
Finally, we have 
\begin{align}
\dfrac{1}{Z}\dfrac{\partial Z}{\partial\psi} &\approx \dfrac{1}{\chi}\left(1+\epsilon\right) \times 2\pi R_0 \frac{\sin2\psi}{\cosh2\mu_0} I_2(\kappa) \\
&= \dfrac{\sin2\hat{\psi}}{\cosh2\mu_0}\dfrac{I_2(\kappa)}{I_0(\kappa)} (1+\epsilon)\\
&= \dfrac{\sin2\hat{\psi}}{\cosh2\mu_0}\dfrac{I_2(\kappa)}{I_0(\kappa)}  + O(\epsilon^2),
\end{align}
where in the last step we distribute the $(1+\epsilon)$ term and neglect all terms of order $\epsilon^2$ since the prefactor $(\cosh 2 \mu_0)^{-1}\sim\epsilon$, which is the small term we are expanding in. 

Armed with this result for $\frac{1}{Z}\frac{\partial Z}{\partial \psi}$, we can find an approximate formula for the maximum likelihood estimator from Eq. \eqref{eq:MLE},
\begin{equation}
	\dfrac{1}{N}\sum_{i=1}^{N}\kappa\sin{(\nu_i-\hat{\psi})}=\dfrac{\sin2\hat{\psi}}{\cosh2\mu_0}\dfrac{I_2(\kappa)}{I_0(\kappa)}.
\end{equation}

Using the identity $\sin(\nu-\hat{\psi})=\sin\nu\cos\hat{\psi}-\cos\nu\sin\hat{\psi}$, we can recast the equation as
\begin{equation}
	\label{eq:MLE_perturbation}
	\left(\dfrac{1}{N}\sum_{i=1}^{N}\sin\nu_i\right)\cos\hat{\psi}-\left(\dfrac{1}{N}\sum_{i=1}^{N}\cos\nu_i\right)\sin\hat{\psi}=\left[\dfrac{1}{\kappa\cosh2\mu_0}\dfrac{I_2(\kappa)}{I_0(\kappa)}\right]\sin2\hat{\psi}.
\end{equation}
The equation is in the form $A\cos\hat{\psi} -B\sin\hat{\psi}=C\sin2\hat{\psi}$. We do not have an analytic solution to this, but we have solved this numerically and find consistent answers with our numerical optimization to find the maximum likelihood estimator.

More importantly, we can use this same perturbation technique to evaluate the integrals in Eq. \eqref{eq:FisherInformation} for the Fisher information. From Eq. \eqref{eq:MLE_perturbation}, we see that in the limit of large $N$, $N^{-1}\sum\sin{(\nu-\psi)}\rightarrow\left\langle\sin{(\nu-\psi)}\right\rangle\sim\left(\cosh2\mu_0\right)^{-1}$. Since the Fisher information contains $\left\langle\sin{(\nu-\psi)}\right\rangle^2$, this term will be proportional to $\left(\cosh2\mu_0\right)^{-2}$, which will be small in our Taylor expansion. Thus, we can ignore it and calculate the Fisher information in a more reduced form:
\begin{equation}
	\label{eq:FI_reduced}
	\mathcal{I}\approx N\kappa^2\left\langle\sin^2{(\nu-\psi)}\right\rangle.
\end{equation}
Expanding like we have done before:
\begin{equation}
	\left\langle\sin^2{(\nu-\psi)}\right\rangle=\dfrac{1}{Z}\int_\gamma\sin^2{(\nu-\psi)}c(\nu)h_{\mu_0}\mathrm{d}\nu\approx(1+\epsilon)\dfrac{R_0}{\chi}\int_\gamma\sin^2{(\nu-\psi)}\exp{[\kappa\cos(\nu-\psi)]}\left[1-\dfrac{1}{2}\left(\dfrac{\cos2\nu}{\cosh2\mu_0}\right)\right]\mathrm{d}\nu.
\end{equation}
Using the substitution $\alpha=\nu-\psi$ again, the identity $\sin^2\alpha=(1-\cos2\alpha)/2$, and common modified Bessel function relations, we can evaluate each of the two terms of the Taylor expansion in the integrand.

\begin{equation}
	\begin{split}
	\left\langle\sin^2{(\nu-\psi)}\right\rangle&\approx\dfrac{R_0}{\chi}(1+\epsilon)\left[\dfrac{1}{2}\left[2\pi I_0(\kappa)-2\pi I_2(\kappa)\right]-\dfrac{2\pi I_2(\kappa)\cos2\psi}{4\cosh2\mu_0}+\dfrac{2\pi\cos2\psi}{8\cosh2\mu_0}[ I_0(\kappa)+ I_4(\kappa)]\right]\\[10pt]
	&=\dfrac{1}{\kappa}\dfrac{I_1(\kappa)}{I_0(\kappa)}+\dfrac{\cos2\psi}{8\kappa\cosh2\mu_0}\left[4\dfrac{I_1(\kappa)I_2(\kappa)}{I_0^2(\kappa)}-\dfrac{\kappa[2I_2(\kappa)-I_0(\kappa)-I_4(\kappa)]}{4I_0(\kappa)}\right]+O(\epsilon^2).
	\end{split}
\end{equation} 
We see now that, in the limit of nearly circular cells, the Fisher information reduces to
\begin{equation}
	\label{eq:FisherInformation_perturbation}
	\mathcal{I}=N\kappa\dfrac{I_1(\kappa)}{I_0(\kappa)}+N\kappa\dfrac{\cos2\psi}{8\cosh2\mu_0}\left[\dfrac{4I_1(\kappa)I_2(\kappa)}{I_0^2(\kappa)}-\dfrac{\kappa[2I_2(\kappa)-I_0(\kappa)-I_4(\kappa)]}{I_0(\kappa)}\right].
\end{equation}
The first term is exactly the Fisher information for a circle \cite{nwogbaga2023physical}, while the second term is the first order correction. This can be rewritten 
\begin{equation}
	\mathcal{I}=N\kappa\left(\dfrac{I_1(\kappa)}{I_0(\kappa)}+\zeta_2\cos2\psi\right),
\end{equation}
where 
\begin{equation}
	\zeta_2 = \dfrac{1}{8\cosh2\mu_0}\left[\dfrac{4I_1(\kappa)I_2(\kappa)}{I_0^2(\kappa)}-\dfrac{\kappa[2I_2(\kappa)-I_0(\kappa)-I_4(\kappa)]}{I_0(\kappa)}\right].
\end{equation}

\subsection{Small $\kappa$ limit Fisher information}
\label{app:FI_smallkappa}

Experimentally, we suspect that the weak-field limit is most relevant. In our previous manuscript studying the physical limits of galvanotaxis on round cells \cite{nwogbaga2023physical}, we discovered that the Fisher information for round cells simplifies to $N\kappa^2/d$, where $d=2,3$ for the dimension (circle versus sphere). Thus, it would nice to derive an expression for the Fisher information in this weak-field limit for an elliptical cell. This requires doing a Taylor expansion in $\kappa$. The concentration $c(\nu)$ is the only function that depends on $\kappa$ and will be perturbed. We want the Fisher information to be of order $\kappa^2$ since it has a prefactor of $\kappa^2$. That would require us to expand the concentration to zeroth order in $\kappa$, meaning $c(\nu)\approx1$. We noticed that at weak fields (small $\kappa$), $\partial_\psi Z\approx0$, meaning that $\left\langle\sin{(\nu-\psi)}\right\rangle\approx0$.  We know this because
\begin{equation}
	\dfrac{1}{Z}\dfrac{\partial Z}{\partial\psi}\approx\ddfrac{\int_\gamma\kappa\sin{(\nu-\psi)}(1+\cdot\cdot\cdot)h_{\mu_0}\mathrm{d}\nu}{\int_\gamma\left(1+\cdot\cdot\cdot\right)h_{\mu_0}\mathrm{d}\nu}.
\end{equation}
where $\cdots$ indicates terms neglected in the limit of small $\kappa$. The numerator integrates to zero since
\begin{equation}
	\int_\gamma\sin{(\nu-\psi)}h_{\mu_0}\mathrm{d}\nu=0.
\end{equation}
This gives us Eq. \eqref{eq:FI_reduced} again for the Fisher information:
\begin{equation}
\mathcal{I}\approx N\kappa^2\left\langle\sin^2{(\nu-\psi)}\right\rangle.
\end{equation}
We only have to evaluate the average $\left\langle\sin^2{(\nu-\psi)}\right\rangle$ to zeroth order in $\kappa$ -- i.e. we evaluate it for a uniform distribution on the ellipse, or $c(\nu) = 1$. This gives us %
\begin{equation}
		\left\langle\sin^2{(\nu-\psi)}\right\rangle=\dfrac{1}{Z}\int_\gamma\sin^2{(\nu-\psi)}c(\nu)h_{\mu_0}\mathrm{d}\nu\approx\ddfrac{\int_\gamma\sin^2{(\nu-\psi)}h_{\mu_0}\mathrm{d}\nu}{\int_\gamma h_{\mu_0}\mathrm{d}\nu}.%
 \end{equation}
 Applying $\sin^2x=(1-\cos2x)/2$:
 \begin{align}
    \left\langle\sin^2{(\nu-\psi)}\right\rangle &=\dfrac{1}{2}-\dfrac{1}{2\int_\gamma h_{\mu_0}\mathrm{d}\nu}\int_\gamma\cos{[2(\nu-\psi)]}h_{\mu_0}\mathrm{d}\nu
    \\[10pt]
    &\equiv\dfrac{1}{2}-(-\xi).
\end{align}
Overall, we see that at small $\kappa$, the Fisher information takes the form
\begin{align}
	\mathcal{I}&\approx N\kappa^2\left(\dfrac{1}{2}+\xi\right),
\end{align} 
where 
\begin{equation}
	\xi=-\frac{1}{2\int_\gamma h_{\mu_0}\mathrm{d}\nu}\int_\gamma\cos{[2(\nu-\psi)]}h_{\mu_0}\mathrm{d}\nu.
\end{equation}
$\xi$ is the anisotropic contribution, i.e. the contribution to the Fisher information from the eccentricity of the cell. The $\psi$ dependence can be factored out of $\xi$. We must evaluate the integral to see this. The integral $\int\cos{[2(\nu-\psi)]}h_{\mu_0}\mathrm{d}\nu$ was evaluated in the interval $[0,2\pi]$. This should be equivalent to integrating in the interval $[\psi-\pi,\psi+\pi]$. We see that
\begin{multline}
\label{eq:xinumerator}
	\int_\gamma\cos{[2(\nu-\psi)]}h_{\mu_0}\mathrm{d}\nu=%
	-\dfrac{a}{3}\cos{2\psi}\left[\left(\cosh\mu_0+\cosh3\mu_0\right)\mathcal{E}\left(\dfrac{1}{\cosh^2{\mu_0}}\right)+2\sinh\mu_0\cosh2\mu_0\mathcal{E}\left(\dfrac{-1}{\sinh^2{\mu_0}}\right)\right]\\[10pt]
	-\dfrac{a}{3}\cos{2\psi}\left[-2\sinh\mu_0\sinh2\mu_0\mathcal{K}\left(\dfrac{1}{\cosh^2{\mu_0}}\right)-2\cosh\mu_0\sinh2\mu_0\mathcal{K}\left(\dfrac{-1}{\sinh^2{\mu_0}}\right)\right],
\end{multline}
where $\mathcal{K}(\cdot)$ and $\mathcal{E}(\cdot)$ are complete elliptical integrals of the first and second kinds, respectively, having the forms
\begin{equation}
	\mathcal{K}\left(\Delta\right)=\int_{0}^{\pi/2}\mathrm{d}\nu\dfrac{1}{\sqrt{1-\Delta\sin^2\nu}},\qquad\mathcal{E}\left(\Delta\right)=\int_{0}^{\pi/2}\mathrm{d}\nu\sqrt{1-\Delta\sin^2\nu}.
\end{equation}
We can also evaluate
\begin{equation}
Z(\kappa = 0) =\int_\gamma h_{\mu_0}\mathrm{d}\nu=\int_\gamma a\sqrt{\dfrac{\cosh{2\mu_0}-\cos{2\nu}}{2}}\mathrm{d}\nu=2a\left[\mathcal{E}\left(\dfrac{1}{\cosh^2{\mu_0}}\right)\cosh\mu_0+\mathcal{E}\left(\dfrac{-1}{\sinh^2{\mu_0}}\right)\sinh\mu_0\right].
\end{equation}
This normalization factor at zero field is simply the perimeter of the ellipse. We can see in Eq. \eqref{eq:xinumerator} that we can extract the $\psi$ dependence:
\begin{equation}
	\xi\equiv\zeta_1(\mu_0)\cos{2\psi}.
\end{equation}
$\zeta_1$ is another constant of aeolotropy, whose value is determined by the elliptical cell radius $\mu_0$, which defines the cell surface. More anisotropic cells have smaller values for $\mu_0$. Fully expanding, $\zeta_1$ appears as
\begin{equation}
	\zeta_1(\mu_0)=\dfrac{1}{12}\dfrac{\left(\cosh\mu_0+\cosh3\mu_0\right)\mathcal{E}\left(\mathfrak{v}_1\right)+2\sinh\mu_0\cosh2\mu_0\mathcal{E}\left(\mathfrak{v}_2\right)
		-2\sinh\mu_0\sinh2\mu_0\mathcal{K}\left(\mathfrak{v}_1\right)-2\cosh\mu_0\sinh2\mu_0\mathcal{K}\left(\mathfrak{v}_2\right)}{\cosh\mu_0\mathcal{E}\left(\mathfrak{v}_1\right)+\sinh\mu_0\mathcal{E}\left(\mathfrak{v}_2\right)},
\end{equation}
where $\mathfrak{v}_1=1/\cosh^2\mu_0$ and $\mathfrak{v}_2=-1/\sinh^2\mu_0$. Now, we can express the Fisher information:
\begin{equation}
	\label{eq:FI_simplifyfull}
	\mathcal{I}\approx N\kappa^2\left(\dfrac{1}{2}+\zeta_1\cos2\psi\right).
\end{equation}

We derived Eq. \eqref{eq:FI_simplifyfull} from assuming the weak-field (small $\kappa$) limit and expanding Eq. \eqref{eq:FisherInformation} from the main text. We can also do a weak-field perturbation from Eq. \eqref{eq:FisherInformation_perturbation} which was derived solely for a nearly circular cell. Expanding each term so that the entire expression is of order $\kappa^2$, we see that 
\begin{equation}
	\mathcal{I}\approx N\kappa\left(\dfrac{\kappa}{2}+\cdot\cdot\cdot\right)+N\kappa\dfrac{\cos2\psi}{8\cosh2\mu_0}(\kappa+\cdot\cdot\cdot)=N\kappa^2\left(\dfrac{1}{2}+\zeta_0\cos2\psi\right),
\end{equation}
where $\zeta_0(\mu_0)=(8\cosh2\mu_0)^{-1}$ is the aeolotropic constant discussed in the main text.

\subsection{Fisher information approximations and aeolotropic constants}
\label{app:FI_approx}

We have been able to show that the Fisher information can be approximated in three different regimes: small $\kappa$, nearly circular cells, and both conditions combined. The approximations are  
\begin{align}
	\mathcal{I}_\mathrm{0}&=N\kappa^2\left(\dfrac{1}{2}+\zeta_0\cos2\psi\right) \; \; \textrm{(small $\kappa$, near-circular)}, \\[10pt]
	\mathcal{I}_\mathrm{1}&=N\kappa^2\left(\dfrac{1}{2}+\zeta_1\cos2\psi\right) \; \; \textrm{(small $\kappa$)}, \\[10pt]
	\mathcal{I}_\mathrm{2}&=N\kappa\left(\dfrac{I_1(\kappa)}{I_0(\kappa)}+\zeta_2\cos2\psi\right) \; \; \textrm{(near-circular)},
\end{align} 
where $\zeta_0$, $\zeta_1$, and $\zeta_2$ are the aeolotropic constants. 
\begin{align}
	\zeta_0&=\dfrac{1}{8}\dfrac{1}{\cosh{2\mu_0}},\\[10pt]
	\zeta_1&=\dfrac{1}{12}\dfrac{\left(\cosh\mu_0+\cosh3\mu_0\right)\mathcal{E}\left(\mathfrak{v}_1\right)+2\sinh\mu_0\cosh2\mu_0\mathcal{E}\left(\mathfrak{v}_2\right)-2\sinh\mu_0\sinh2\mu_0\mathcal{K}\left(\mathfrak{v}_1\right)-2\cosh\mu_0\sinh2\mu_0\mathcal{K}\left(\mathfrak{v}_2\right)}{\cosh\mu_0\mathcal{E}\left(\mathfrak{v}_1\right)+\sinh\mu_0\mathcal{E}\left(\mathfrak{v}_2\right)},\\
	\zeta_2 &= \dfrac{1}{8\cosh2\mu_0}\left[\dfrac{4I_1(\kappa)I_2(\kappa)}{I_0^2(\kappa)}-\dfrac{\kappa[2I_2(\kappa)-I_0(\kappa)-I_4(\kappa)]}{I_0(\kappa)}\right].
\end{align}
$\zeta_0$ and $\zeta_2$ can easily be written in a more intuitive manner by representing them in terms of the aspect ratio $\lambda$. Recall that $\coth\mu_0=\lambda$. This means that
\begin{equation}
	\cosh2\mu_0=\cosh[2 \mathrm{arcoth}(\lambda)]=\dfrac{\lambda^2+1}{\lambda^2-1}.
\end{equation}
We can now rewrite the constants $\zeta_0$ and $\zeta_2$
\begin{align}
	\zeta_0&=\dfrac{1}{8}\dfrac{\lambda^2-1}{\lambda^2+1},\\[10pt]
	\zeta_2&=\zeta_0\left[\dfrac{4I_1(\kappa)I_2(\kappa)}{I_0^2(\kappa)}-\dfrac{\kappa[2I_2(\kappa)-I_0(\kappa)-I_4(\kappa)]}{I_0(\kappa)}\right]\equiv\zeta_0A(\kappa).
\end{align}
The Fisher information consists of two components: the circular portion and the anisotropic portion. The circular portion represents the Fisher information for a circular cell. In contrast, the anisotropic portion accounts for the cell's eccentricity and is governed by aeolotropic constants. These aeolotropic constants should vanish when the Fisher information pertains to a circular cell. It is trivial to check that $\zeta_0=\zeta_2=0$ when $\lambda=1$. $\zeta_1$ is less obvious. To see this, note that when we have a circle, all the elliptical integrals evaluate to $\pi/2$. This is because in the limit of a circle, $\lambda\rightarrow1$ implies that $\mu_0\rightarrow\infty$. Then, by multiplying $\zeta_1$ by $1=a/a$, the numerator of $\zeta_1$ becomes 
\begin{equation}
	a\pi\left(\dfrac{1}{2}\cosh\mu_0+\dfrac{1}{2}\cosh3\mu_0-\sinh\mu_0\sinh2\mu_0-(\sinh2\mu_0\cosh\mu_0-\sinh\mu_0\cosh2\mu_0)\right).
\end{equation}
We can use the identity $\sinh(x-y)=\sinh x\cosh y-\sinh y\cosh x$ to simplify the above expression
\begin{equation}
	a\pi\left(\dfrac{1}{2}\cosh\mu_0+\dfrac{1}{2}\cosh3\mu_0-\sinh\mu_0\sinh2\mu_0-\sinh\mu_0\right).
\end{equation}
Similarly, we can use $\cosh{(x\pm y)}=\cosh x\cosh y\pm\sinh x\sinh y$ to simplify our expression in the numerator of $\zeta_1$ further:
\begin{equation}
	a\pi\left(\cosh\mu_0-\sinh\mu_0\right)\equiv\pi(R_1-R_2). 
\end{equation}
Since $R_1=R_2=R_0$ for a circle, $\zeta_1=0$ and we have our desired limit.

We just showed that the aeolotropic constants have a minimum value $\zeta_\mathrm{min}=0$ for a circle, which occurs when $\lambda=1$. What are the values when we have an infinitely eccentric cell (1-dimensional line)? Do they have maxima? An infinitely eccentric cell occurs when $\mu_0\rightarrow0^+$ ($\lambda\rightarrow\infty$). In this limit, it is easy to see that $\zeta_0\rightarrow1/8$ and $\zeta_2\rightarrow A(\kappa)/8$. Checking for $\zeta_1$ is a bit tougher. We see that in this limit, $\cosh\mu_0=1$, $\sinh\mu_0=0$, $\mathcal{K}\left(\mathfrak{v}_2\right)=0$, $\mathcal{E}\left(\mathfrak{v}_1\right)=1$. $\mathcal{K}\left(\mathfrak{v}_1\right)$ becomes a complex number, where $\Re{[\mathcal{K}\left(\mathfrak{v}_1\right)]}\rightarrow\infty$ and $\Im{[\mathcal{K}\left(\mathfrak{v}_1\right)]}=\pi$. Meanwhile, $\mathcal{E}\left(\mathfrak{v}_2\right)\rightarrow\infty$. Fortunately, $\sinh\mu_0$ decreases at a faster rate than $\mathcal{K}(\cdot)$ and $\mathcal{E}(\cdot)$ can blow up. That means that $\zeta_1\rightarrow(2+0-0-0)/12=1/6$. Armed with that information, we can summarize our findings:
\begin{align}
	&\lim\limits_{\mu_0\rightarrow\infty}\zeta_0,\zeta_1,\zeta_2\equiv\zeta_\mathrm{min}=0\quad\textrm{(circle)},\\[10pt]
	&\lim\limits_{\mu_0\rightarrow0^+}\zeta_0=\dfrac{1}{8},\quad\lim\limits_{\mu_0\rightarrow0^+}\zeta_1=\dfrac{1}{6},\quad\lim\limits_{\mu_0\rightarrow0^+}\zeta_2=\dfrac{A(\kappa)}{8}\quad\textrm{(line)}.
\end{align}

\section{Normal approximation and fitting $\gamma$}
\label{app:directionality}

\begin{figure}[htbp]
	\centering
	\includegraphics[width=0.8\linewidth]{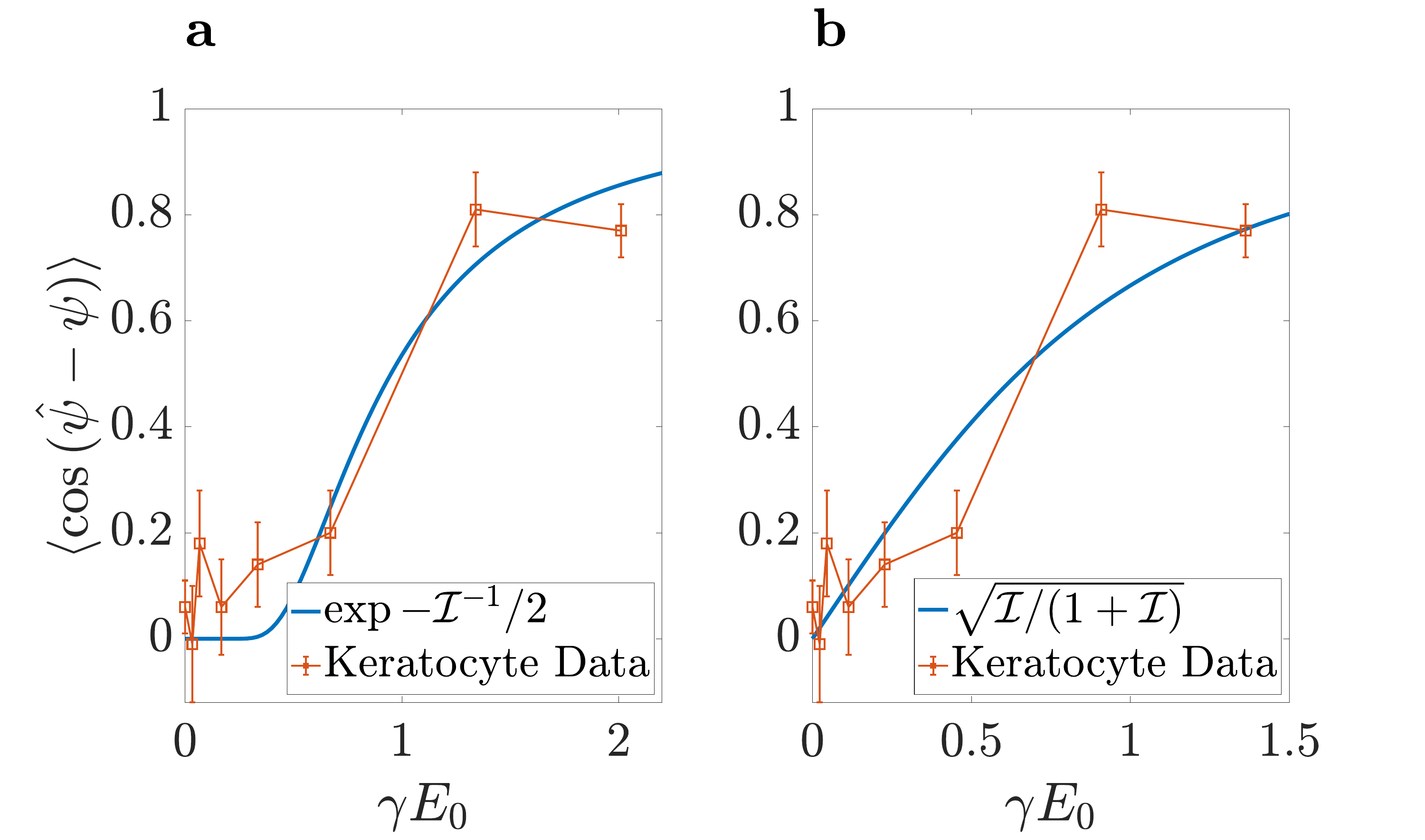}
	\caption{Plots fitting Eq. \eqref{eq:cos_directionality} directionality to keratocyte data where $\mathcal{I}=0.8(\gamma E_0)^2$. (a) Fit with $e^{\mathcal{I}^{-1}/2}$, yielding a fit value of $\gamma=3.4\times10^{-3}$ mm/mV. (b) Fit with $\sqrt{\mathcal{I}/(\mathcal{I}+1)}$, yielding a fit value of $\gamma=2.3\times10^{-3}$ mm/mV. }
	\label{fig:keratocytefitdirectionality}
\end{figure}

In the main text, we claim that the directionality, the measure of how well cells follow electric fields \cite{nwogbaga2023coupling,nwogbaga2023physical}, is $\langle\cos{(\hat\psi-\psi)}\rangle\approx e^{-\mathcal{I}^{-1}/2}$. This is an alternate formula to $\langle\cos{(\hat\psi-\psi)}\rangle\approx\sqrt{\mathcal{I}/(1+\mathcal{I})}$, which we used in \cite{nwogbaga2023physical}. We show fits to both forms in Fig. \ref{fig:keratocytefitdirectionality}. This alternate directionality formula was derived by taking advantage of the property that maximum likelihood estimators are asymptotically normal in the limit of large sample sizes \cite{baum1966statistical,lehmann1999elements}. Thus, for a sufficiently large sample of sensors $N$, even the periodic estimator $\hat\psi\xrightarrow{\text{d}}\mathcal{N}(\psi,\mathcal{I}^{-1})$. (Note that in the literature, the Fisher information per observation $\mathcal{I}_N$ is often used, making the limiting variance $(N\mathcal{I}_N)^{-1}=\mathcal{I}^{-1}$). That means the density for $\hat\psi$ can be approximated by a Gaussian distribution
\begin{equation}
	f(\hat\psi)\approx\sqrt{\dfrac{\mathcal{I}}{2\pi}}\exp{\left[-\dfrac{\mathcal{I}}{2}(\hat\psi-\psi)^2\right]}.
\end{equation} 
With this probability density function, $\langle\cos{(\hat\psi-\psi)}\rangle$ can be computed 
\begin{equation}
	\label{eq:cos_directionality}
	\langle\cos{(\hat\psi-\psi)}\rangle\approx\int_{-\infty}^{\infty}\cos{(\hat\psi-\psi)}f(\hat\psi)\mathrm{d}\hat\psi=e^{-\mathcal{I}^{-1}/2}.
\end{equation}

We derive the Fisher information limit appropriate for keratocytes in Eq. \eqref{eq:keratocyte_FI} in the main text. We use this result to fit the directionality $\langle\cos{(\hat\psi-\psi)}\rangle$ to keratocyte experimental data from \cite{sun2013keratocyte} to determine a reasonable value for $\gamma$, the characteristic electric field strength. Data of keratocyte directionality as a function of field strength was fitted using Eq. \eqref{eq:keratocyte_FI} and plugging it into Eq. \eqref{eq:cos_directionality}. This fitting process produced $\gamma=3.4\times10^{-3}$ mm/mV (Fig. \ref{fig:keratocytefitdirectionality}a). We see in Fig. \ref{fig:keratocytefitdirectionality} that the normal-approximation assumption is a slightly better fit to the experimental data, though we are not confident that this data can really discriminate between the two models. The difference in $\gamma$ in fitting to these two models is relatively small. However, there would be minor quantitative changes if we chose the alternate value of $\gamma = 2.3 \times 10^{-3}$ mm/mV. This sets the scale of electric field the cell can sense, and would make the transition in trends in Fig. \ref{fig:keratocyte} in the main text occur at a higher field strength and the magnitude of the variance decrease more slowly as the electric field strength is increased. 

\section{Circular variance derived from MLE for keratocytes as a function of sensor number $N$}
\label{app:keratocytefit}

\begin{figure}[htbp]
	\centering
	\includegraphics[width=0.6\linewidth]{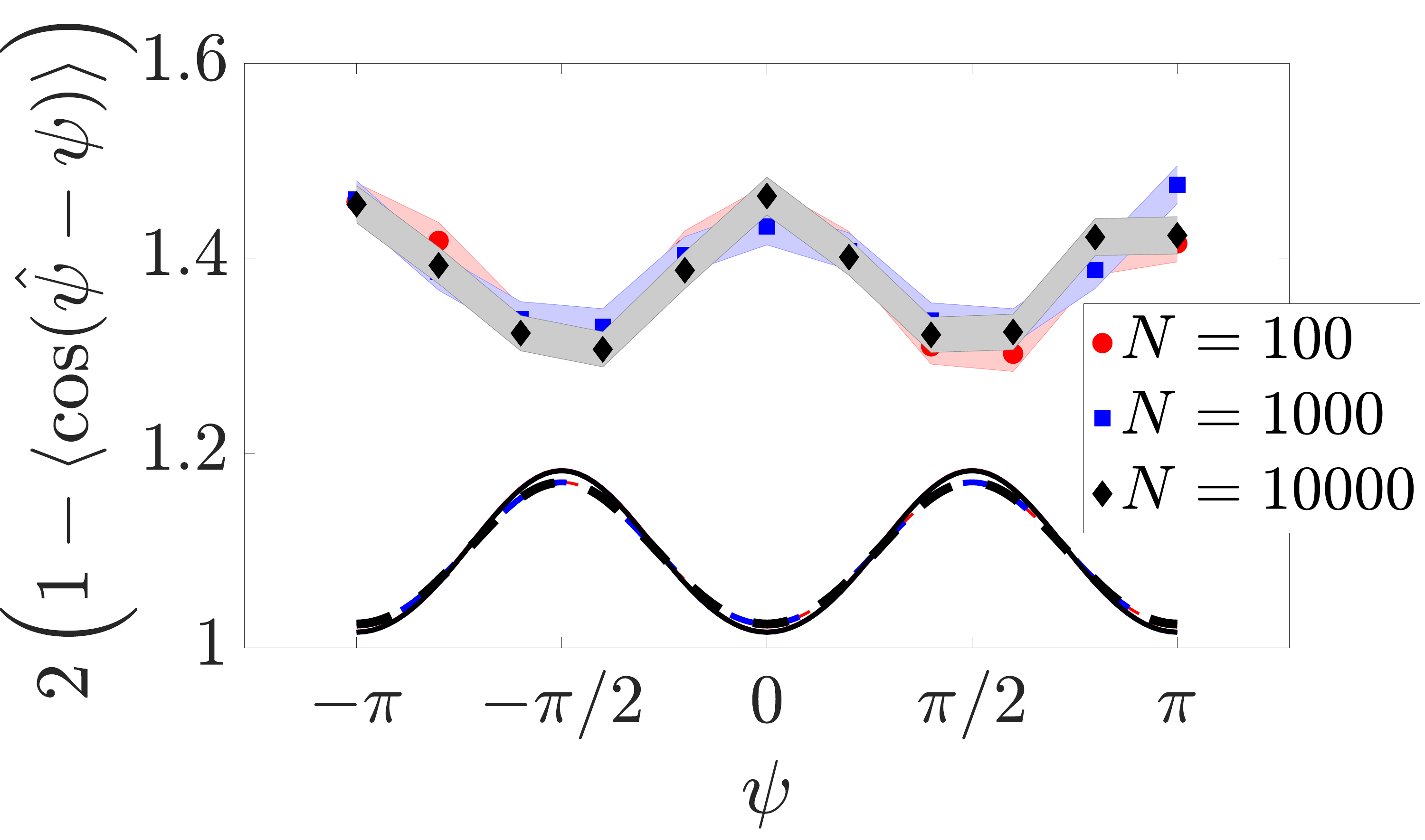}
	\caption{Circular variance plots for keratocytes calculated using MLE. $\gamma=3.4\times10^{-3}$ mm/mV is kept constant. Simulated for 5000 cells at a field strength $E_0=150$ mV/mm. Varied for $N=100$ $(\kappa\approx5\times10^{-2})$, $N=1000$ $(\kappa\approx1.5\times10^{-2})$, and $N=10000$ $(\kappa\approx5\times10^{-3})$. Solid lines is the lower bound from Eq. \eqref{eq:circularvariance}. Dashed lines are the lower bound from Eq. \eqref{eq:FI_simplifyweak}. Shaded region are error bars for standard error of the mean.}
	\label{fig:keratocytefit}
\end{figure}

We found that the Fisher information does not depend on $\beta$ and $N$ independently, but only on the combination $\gamma=N\beta^2/2$. However, because in the limit of weak fields, the maximum likelihood estimator's variance is above the Cramer-Rao bound, it is possible that the MLE variance depends separately on $N$ and $\beta$. Here, we show how the circular variance changes while the number of sensors $N$ changes, while keeping $\gamma$ constant -- and thus keeping the magnitude of the Fisher information constant. If we fix $\gamma=3.4\times10^{-3}$ mm/mV, then the behavior of the circular variance is unaffected by changing $N$ (Fig. \ref{fig:keratocytefit}). According to the simulations, the circular variance remains maximal when the electric field is parallel to the cell's long axis and minimal if the field is parallel to the cell's short axis.

\newpage
\section*{References}

\end{document}